\documentclass[epj-spec]{svjour}

\usepackage{graphicx} % including PostScript
\usepackage{amsmath}
\usepackage{amssymb} % e.g. \gtrsim
\usepackage{bbm} % BlackBoeard letters
\usepackage{mathrsfs}	% e.g. nice Lagrange-L with \mathscr{L}
\usepackage{fleqn} % equations are not centered
\usepackage[small,loose]{subfigure}  % subfigures with (a), (b) etc., ... and own caption
\usepackage{paralist}
\usepackage{pifont}
\usepackage{picins} 
\usepackage{pst-all} % equations are not centered
\usepackage{framed}
\newrgbcolor{shadecolor}{0.8 0.85 1}
\definecolor{darkgreen}{rgb}{0.0, 0.6, 0.2}
\newrgbcolor{darkgreen}{0.0 0.6 0.2}
\definecolor{purple}{rgb}{0.75, 0.2, 0.75}
\newrgbcolor{purple}{0.75 0.2 0.75}
\newgray{gray90}{0.9}
\newgray{gray80}{0.8}
\newgray{gray70}{0.7}

\newcommand{\CenterObject}[1]{\ensuremath{\vcenter{\hbox{#1}}}}
\newcommand{\CenterEps}[2][1]{\ensuremath{\vcenter{%
      \hbox{\includegraphics[scale=#1]{#2.eps}}}}} 
\newcommand{\BmL}{\ensuremath{B\!-\!L} }

\newcommand{\E}[1]{\ensuremath{\mathrm{E}_{#1}}} % e.g. \E{8}
\newcommand{\G}[1]{\ensuremath{\mathrm{G}_{#1}}}
\newcommand{\SO}[1]{\ensuremath{\mathrm{SO}(#1)}}
\newcommand{\SU}[1]{\ensuremath{\mathrm{SU}(#1)}}
\newcommand{\U}[1]{\ensuremath{\mathrm{U}(#1)}}
\newcommand{\Z}[1]{\ensuremath{\mathbbm{Z}_{#1}}} % Z_N ->\Z{N}

\DeclareMathOperator{\re}{Re}

\begin{document}
\title{{}\\[-1.5cm]
{\normalsize\normalfont{\vspace*{1cm}\hfill TUM-HEP 690/08}}\\
From strings to the MSSM}
\subtitle{}
\author{Hans Peter Nilles\inst{1}\fnmsep\thanks{\email{nilles@th.physik.uni-bonn.de}} 
\and 
Sa\'ul Ramos-S\'anchez\inst{1}\fnmsep\thanks{\email{ramos@th.physik.uni-bonn.de}}
\and 
Michael Ratz\inst{2}\fnmsep\thanks{\email{mratz@ph.tum.de}}
\and 
Patrick K.S. Vaudrevange\inst{1}\fnmsep\thanks{\email{patrick@th.physik.uni-bonn.de}}  }
\institute{Physikalisches Institut der Universit\"at Bonn, Nussallee 12, 53115
Bonn, Germany 
\and 
Physik Department T30, Technische Universit\"at M\"unchen, James-Franck-Strasse,
85748 Garching, Germany}
\abstract{We review recent progress in embedding the supersymmetric standard
model into string theory. We discuss how, with the incorporation of certain
aspects of grand unification, a search strategy can be developed that allows to
efficiently find rather large numbers of promising string vacua. Global
string-derived models with the following features are discussed: (i) exact MSSM
spectrum below the unification scale; (ii) R parity; (iii) hierarchical Yukawa
couplings with non-trivial mixing; (iv) solution to the $\mu$ problem; (v)
see-saw suppressed neutrino masses. 
}
\maketitle

\section{Introduction}

Attempts to relate superstring theory to models of elementary particle physics
do not yet provide us with a clear solution.
While string theory contains everything which is needed to describe the real
world, i.e.\ gravity, gauge interactions and chiral matter, the `details' appear
not quite to fit observation: string theory predicts 10 rather than 4 space-time
dimensions, $N=4$ or $N=8$ instead of $N=0$ (or $N=1$) supersymmetry and gauge
groups as large as $\E8\times\E8$ or \SO{32} instead of the standard model
(SM) gauge group,
\begin{equation}
 G_\mathrm{SM}~=~\SU3_\mathrm{C}\times\SU2_\mathrm{L}\times\U1_Y\;.
\end{equation}
% $G_\mathrm{SM}=\SU3_\mathrm{C}\times\SU2_\mathrm{L}\times\U1_Y$.  
It is generally expected that these discrepancies can be eliminated by
compactifying six spatial dimensions on suitable manifolds. Different shapes of
compact space will lead to different phenomenology. The standard model fields
will then be interpreted as (almost) massless vibration modes of internal space.
One might think that one would just have to `dial' the right compact space to
obtain the standard model.  Unfortunately this is not the case: since string
theory is rather restrictive, even when one allows to choose the internal
manifold at will it is non-trivial to obtain something that looks like the
standard model at low energies. Stringy consistency conditions often enforce the
presence of unwanted states in the low-energy spectrum and/or lead to
unrealistic couplings. How can we find realistic string compactifications then?
A blind scan is hopeless since string theory equations have many solutions,
i.e.\ there is a huge number of 4D vacua (the so-called string-theory
landscape), which cannot be analyzed systematically. To obtain predictive string
models one therefore has to develop a suitable strategy.

In this paper we explain how low-energy observations allow us to draw some
conclusions on the generic properties of compact space. We know that the
standard model gauge group is unbroken above the electroweak scale and that most
of the Yukawa couplings are hierarchically small. We will argue that these facts
allow us to conclude that we live close to a special (`symmetry-enhanced') point
in  moduli space. Furthermore, the successful aspect of grand unified theories
(GUTs) indicate that \SO{10} structures represent an important ingredient of the
stringy completion of the SM. From these considerations we infer that the heterotic string
\cite{Gross:1985fr,Gross:1985rr} compactified on an orbifold
\cite{Dixon:1985jw,Dixon:1986jc,Ibanez:1986tp,Ibanez:1987sn,Casas:1987us,Casas:1988hb,Font:1988tp,Font:1988nc,Font:1988mm,Font:1989aj}
can be a good starting point for the task of embedding the SM into string
theory. First, the orbifold point denotes a symmetry-enhanced point in moduli space
(away from it the gauge symmetry gets reduced). Second, the gauge group \E8, which is
specific to the heterotic string, contains an \SO{10} subgroup as well as the
16-dimensional \SO{10} spinor in the coset $\E8/\SO{10}$, which unifies one
family of quarks and leptons. Orbifold compactifications have a long history
(see \cite{Quevedo:1996sv,Bailin:1999nk} for earlier reviews); the focus of this
paper is to review recent developments which allow, by incorporating certain
aspects of grand unification, to construct phenomenologically attractive string
models. (For reviews of alternative approaches leading to interesting models see
e.g.\ \cite{Blumenhagen:2005mu,Blumenhagen:2006ci,Cleaver:2007ek}.) In the
emerging scheme, low-energy supersymmetry plays a key role.

\section{Supersymmetry and grand unified structures}

As is well known, the concept of grand unification allows to simplify the
observed matter content of the standard model. One generation of standard model
matter, transforming as 
\begin{equation}\label{eq:SMgeneration}
\text{generation}~=~
        (\boldsymbol{3},\boldsymbol{2})_{1/6}+
        (\overline{\boldsymbol{3}},\boldsymbol{1})_{-2/3}+
        (\overline{\boldsymbol{3}},\boldsymbol{1})_{1/3}+
        (\boldsymbol{1},\boldsymbol{2})_{-1/2}+
        (\boldsymbol{1},\boldsymbol{1})_{1}
\end{equation}
under $G_\mathrm{SM}$, can be combined into two \SU5 representations
\cite{Georgi:1974sy},
\begin{subequations}\label{eq:SU5matter}
\begin{eqnarray} 
 \boldsymbol{10}
 & = &  (\boldsymbol{3},\boldsymbol{2})_{1/6}\oplus
        (\overline{\boldsymbol{3}},\boldsymbol{1})_{-2/3}\oplus
        (\boldsymbol{1},\boldsymbol{1})_{1}\;,\\
 \overline{\boldsymbol{5}}		
 & = &
        (\overline{\boldsymbol{3}},\boldsymbol{1})_{1/3}\oplus
        (\boldsymbol{1},\boldsymbol{2})_{-1/2}\;.
\end{eqnarray}
\end{subequations}
These two irreducible representations, together with the right-handed neutrino,
form a 16-dimensional spinor representation of \SO{10} ($\boldsymbol{16}$-plet)
\cite{Georgi:1975qb,Fritzsch:1974nn}, i.e.\
\begin{eqnarray}
\boldsymbol{16}
 & = & \boldsymbol{10}\oplus\overline{\boldsymbol{5}}\oplus\boldsymbol{1}
 \nonumber\\
 & = &
(\boldsymbol{3}, \boldsymbol{2})_{1/6} \oplus
(\boldsymbol{\overline{3}}, \boldsymbol{1})_{-2/3}\oplus
(\boldsymbol{\overline{3}}, \boldsymbol{1})_{1/3}\oplus
(\boldsymbol{1}, \boldsymbol{2})_{-1/2} \oplus
(\boldsymbol{1}, \boldsymbol{1})_{1}\oplus
(\boldsymbol{1}, \boldsymbol{1})_{0}
 \;.
\end{eqnarray}
These facts represent some of the most compelling evidence for grand
unification. 

The paradigm of grand unified theories (GUTs) gets supported by the observation
that gauge couplings unify in the minimal supersymmetric extension of the
standard model (MSSM) \cite{Amaldi:1991cn} at the GUT scale
$M_\mathrm{GUT}=\text{few}\times10^{16}\,\mathrm{GeV}$. This scale seems to play
a role also in neutrino physics, where the see-saw \cite{Minkowski:1977sc}
appears to be the most plausible explanation for the smallness of neutrino
masses. That is, $(\text{weak scale})^2/M_\mathrm{GUT}$ gives roughly the right
scale for the observed mass squared differences. Together with other indirect
evidence for supersymmetry, such as a compelling particle physics candidate for
the observed cold dark matter and experimental hints for a light Higgs boson,
this leads to the following popular picture for physics beyond the standard
model: above a scale of the order TeV, the world becomes supersymmetric. The
particle content of the supersymmetric standard model yields an adequate
description up to $M_\mathrm{GUT}$, where gauge couplings meet. At this scale,
new physics appears.

In the scheme of traditional grand unification, this would be a gauge theory
based on a unified (GUT) group. These so-called SUSY GUTs are very popular,
mostly because of the following nice properties:
\begin{dingautolist}{"0C0}
 \item unified multiplets, in particular spinors of \SO{10};
 \item gauge coupling unification;
 \item GUTs yield a reasonable prediction for the see-saw scale;
 \item Yukawa unification, i.e.\ the $\tau$ and $b$ masses enjoy approximate
 unification at the GUT scale, and for suitable $\tan\beta$ also the top mass
 may be unified.
\end{dingautolist}
Arguably, it is hard to believe that these relations are just accidents.
However, the scheme of 4D grand unification exhibits certain problematic
features:
\begin{dingautolist}{"0CA}
 \item the doublet-triplet splitting problem: in a four-dimensional GUT
 theory, the particle content has to respect the GUT symmetry. This implies, in
 particular, to the Higgs fields. However, the smallest \SO{10} representation
 containing the MSSM Higgs doublets (or the SM Higgs) is the
 $\boldsymbol{10}$-plet, which decomposes as 
 \begin{equation}
  \boldsymbol{10}~=~
  (\boldsymbol{1},\boldsymbol{2})_{1/2}
  \oplus
  (\boldsymbol{1},\boldsymbol{2})_{-1/2}
  \oplus
  (\boldsymbol{3},\boldsymbol{1})_{-1/3}
  \oplus
  (\overline{\boldsymbol{3}},\boldsymbol{1})_{1/3}
 \end{equation}
 under $\SO{10}\to G_\mathrm{SM}$. That is, requiring the existence of Higgs
 doublets $(\boldsymbol{1},\boldsymbol{2})_{\pm1/2}$ leads necessarily also to
 color triplets. However, there exist rather uncomfortable lower bounds on the
 mass of these triplets; for instance, in the context of \SO{10} SUSY GUTs, a
 lower bound of about the Planck scale has been
 reported~\cite{Dermisek:2000hr}.  Although this problem may be solved
 \cite{Dimopoulos:1981xm,Masiero:1982fe,Babu:1993we}, the complexity of the
 known solutions casts some shadow on the scheme of 4D grand unification.
 \item While third generation fermion masses seem to comply with grand
 unification \cite{Buras:1977yy}, the GUT fermion mass relations are challenged
 by observation. (There is also a tension between precision unification of the
 third family masses and FCNC constraints \cite{Altmannshofer:2008vr}.)
 \item Breaking of the GUT symmetry requires Higgs fields in large
 representations. It is hard, if not impossible, to get these out of string
 theory, which is the most promising candidate for the description of all
 forces (cf.\ \cite{Dienes:1996yh}).
\end{dingautolist}
These are perhaps the greatest problems of the traditional scheme of grand
unification. This raises the question whether one may modify the scheme in such
a way that \ding{"0C0}--\ding{"0C3} are retained while \ding{"0CA}--\ding{"0CC}
are avoided.

The answer to this question is affirmative: extra dimensions (with size of the
order of $M_\mathrm{GUT}^{-1}$) allow to solve these problems.  In fact, it has
been pointed out in the context of string theory  that such schemes allow for
GUT symmetry breaking without the need for large representations and for
successful doublet triplet splitting \cite{Witten:1985xc,Breit:1985ud,Ibanez:1987sn}.
Simplified versions of the stringy mechanism have been discussed in the context
of what is known as `orbifold GUTs'
\cite{Kawamura:1999nj,Kawamura:2000ev,Altarelli:2001qj,%
Hall:2001pg,Hebecker:2001wq,Asaka:2001eh,Hall:2001xr,Burdman:2002se} (for a review, see e.g.\
\cite{Quiros:2003gg}). 
From these one gains a geometric intuition on certain lower-dimensional building
blocks which facilitate the construction of promising models exhibiting the
appealing features of GUTs while avoiding most of their problems (for a recent
review pursuing a bottom-up approach going from the SM via GUTs and orbifold
GUTs to strings see \cite{Ratz:2007my}).  In what follows, we proceed as
follows: we first review the heterotic string compactified on  orbifolds,
introducing the concepts of `gauge group topographies' and `local GUTs'. We will
show how they can be used to define a search strategy for obtaining promising
models. Finally, we comment on how orbifold GUTs can be derived from string
orbifolds.

\section{Orbifold compactifications}

Having discussed the virtues and problems of grand unification, we continue by
pursuing a more `top-down' approach. We aim at embedding the standard model into string
theory. In our task, we take the positive features of SUSY GUTs
(\ding{"0C0}--\ding{"0C3}) seriously and therefore seek for string models
where they emerge. Grand unified structures seem to require the \SO{10} gauge
group as well as $\boldsymbol{16}$-plets. 
% (As already discussed, deriving 4D GUTs from string theory appears not to be
% promising because it seems difficult, if not impossible, to get the required
% large Higgs representations.) 
Among the known perturbative string theories, only the heterotic string does
possess these ingredients. Hence, we will concentrate on compactifications of the
heterotic string. We also saw that (an approximate) $N=1$ supersymmetry is
crucial. The requirement of $N=1$ supersymmetry restricts the compact space: it
has to be such that the holonomy group fits into an \SU3.\footnote{Note that if 
the holonomy group fits additionally into an $\SU2\subset\SU3$, one retains 
at least $N=2$ supersymmetry.} In the case of a
smooth compactification manifold this means that the 6D space has to be of the
Calabi-Yau type \cite{Candelas:1985en}. But string theory is also well-defined
and regular on non-smooth compactifications, in particular on orbifolds
\cite{Dixon:1985jw,Dixon:1986jc}, which turn out to be easier to deal with. So
our strategy will be to first focus on heterotic orbifolds; later, in
section~\ref{sec:BlowUp}, we will comment on the relation between these
constructions and Calabi-Yau compactifications.

\begin{figure}[h]
\centerline{
\CenterObject{\psset{unit=1cm}
\begin{pspicture}(-3.5,-0.1)(2.4,2.1)
   	\pspolygon[fillstyle=solid,fillcolor=gray90,linecolor=gray80]
		(0,0)(2.309,0)(-1.115,2)(-3.464,2)
   	\psline[linewidth=0.8mm,linecolor=red]{*->}(0,0)(-3.464,2)
   	\psline[linewidth=0.8mm,linecolor=red]{*->}(0,0)(2.309,0)
	\rput[t](2.2,-0.1){\small {\red $\re z_1$}}
	\psdot*[dotstyle=o,dotsize=3mm](0,0)
	\psdot*[dotstyle=*,dotsize=2mm](0,0)
\end{pspicture}
\psset{unit=1cm}}
$\times$
\CenterObject{\psset{unit=1.2cm}
\begin{pspicture}(-1.2,-0.1)(2,1.8)
	\pspolygon[fillstyle=solid,fillcolor=gray90,linecolor=gray80]
		(0,0)(2,0)(1,1.73205)(-1,1.73205)(0,0)
	\psline[linewidth=0.8mm,linecolor=darkgreen]{*->}(0,0)(2,0)
	\psline[linewidth=0.8mm,linecolor=darkgreen]{*->}(0,0)(-1,1.73205)
	\rput[t](1.8,-0.1){\small {\darkgreen $\re z_2$}}
	\psdot*[dotstyle=o,dotsize=3mm](0,0)
 	\psdot*[dotstyle=*,dotsize=2mm](0,0)
	\psdot*[dotstyle=o,dotsize=3mm](1,0.5)
	\psdot*[dotstyle=*,dotsize=2mm](1,0.5)
	\psdot*[dotstyle=o,dotsize=3mm](0.0669873, 1.11603)
	\psdot*[dotstyle=*,dotsize=2mm](0.0669873, 1.11603)
\end{pspicture}\psset{unit=1cm}}
$\times$
\CenterObject{\psset{unit=1.2cm}
\begin{pspicture}(-0.6,-0.1)(2.1,2.1)
	\pspolygon[fillstyle=solid,fillcolor=gray90,linecolor=gray80]
		(0,0)(2,0)(2,2)(0,2)(0,0)
	\psline[linewidth=0.8mm,linecolor=blue]{*->}(0,0)(2,0)
	\rput[t](1.8,-0.1){\small {\blue $\re z_3$}}
	\psline[linewidth=0.8mm,linecolor=blue]{*->}(0,0)(0,2)
	\psdot*[dotstyle=o,dotsize=3mm](0,0)
 	\psdot*[dotstyle=*,dotsize=2mm](0,0)
	\psdot*[dotstyle=o,dotsize=3mm](1,0)
	\psdot*[dotstyle=*,dotsize=2mm](1,0)
	\psdot*[dotstyle=o,dotsize=3mm](1,1)
	\psdot*[dotstyle=*,dotsize=2mm](1,1)
	\psdot*[dotstyle=o,dotsize=3mm](0,1)
	\psdot*[dotstyle=*,dotsize=2mm](0,1)
\end{pspicture}\psset{unit=1cm}}
}
\caption{$\Lambda_{\G2\times\SU3\times\SO4}$ and \Z6 fixed points.}
\label{fig:LambdaG2xSU3xSO4}
\end{figure}
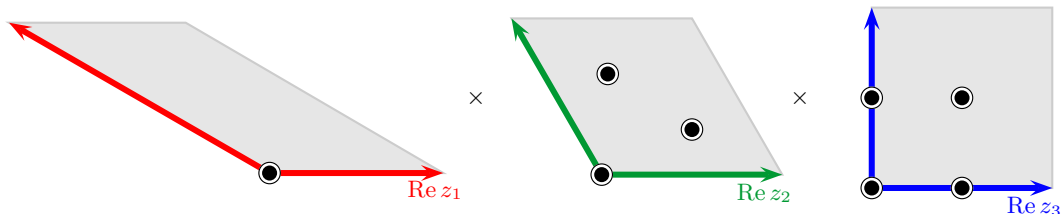

Rather than discussing orbifolds in general, we will focus on a specific example
-- the $\Z6-\mathrm{II}=\Z3\times\Z2$ orbifold. Here, in a first step, six
dimensions get compactified on a torus $\mathbbm{T}^6$ that enjoys a
$\Z3\times\Z2$ discrete symmetry. One example for such a torus is
\begin{equation}
 \mathbbm{T}^6~=~\mathbbm{T}^2_{\G2}\times\mathbbm{T}^2_{\SU3}
 \times\mathbbm{T}^2_{\SO4}\;.
\end{equation}
The subscripts represent the Lie algebra fixing the geometrical relations
of the lattice vectors defining the corresponding torus. For instance,
$\mathbbm{T}^2_{\G2}$ emerges as $\mathbbm{R}^2/\Lambda_{\G2}$, that is, two points
of the two-dimensional plane are identified if they differ by an integer linear
combination of lattice vectors whereby the basis vectors enjoy the same
geometrical relations as the roots of $\G2$, i.e.\ enclose $150^\circ$ and have a
length ratio $1/\sqrt{3}$ (cf.\ figure \ref{fig:LambdaG2xSU3xSO4}). 

The orbifold emerges by modding out the discrete symmetry of the torus,
\begin{equation}
 \text{orbifold}~=~\mathbbm{T}^6/\mathbbm{Z}_6\;,
\end{equation}
where the \Z6 operation $\theta$ acts as a simultaneous rotation by $60^\circ$,
$120^\circ$ and $180^\circ$ on the \G2, \SU3 and \SO4 tori, respectively. This
action is not free, i.e.\ there appear fixed points, which get mapped onto
themselves under $\theta$. It is this orbifold action that breaks $N=4$
supersymmetry in 4D, which one would obtain from a torus compactification, down
to $N=1$. 

Apart from the geometrical properties discussed above, to build an orbifold
model one has to specify the gauge embedding. That is, the geometrical operation
$\theta$ is to be associated to an operation in the $\E8\times\E8$
gauge degrees of freedom. In the bosonic formulation we are using here, this
action is encoded in a 16-dimensional vector $V$, called the shift vector.
This breaks the gauge symmetry from $\E8\times\E8$ to a subgroup, as we will
discuss shortly.

In the heterotic string compactified on an orbifold, there are two types of
closed strings. Firstly, there are ordinary closed strings which are free to
move in the ten-dimensional bulk, called untwisted strings. In addition, there
appear new closed strings, which only close after the action of \Z6 and are
attached to the fixed points. These strings are called twisted strings.  Hence,
we distinguish between different sectors, according to the rotation $\theta^k$
necessary to close the string. Untwisted strings ($k=0$) comprise the untwisted
sectors, denoted by $U_i$, where $i=1,2,3$ refers to the three two-dimensional
planes of the compact space. Twisted strings ($k=1,\ldots,5$) give rise to the
twisted sectors $T_k$.

\subsection{Simple ways of envisaging orbifolds}

Many important aspects of 6D orbifolds can be understood from considering lower-dimensional
versions. We start with $\mathbbm{T}^2/\mathbbm{Z}_2$, which emerges by dividing
the torus $\mathbbm{T}^2$ by a point reflection symmetry. The torus is defined
by 2 (linearly independent) lattice vectors $e_1$ and $e_2$ spanning the
fundamental domain. The $\Z2$ then acts as a reflection (or, equivalently,
$180^\circ$ rotation) about an arbitrary lattice node which one could call
`origin'. Certain points are mapped under the orbifold action onto themselves
(up to lattice translations); these are the orbifold fixed points.
After identifying points in the fundamental domain of the torus that are
related by the \Z2 orbifold action, one arrives at the fundamental domain of the
orbifold. By folding the fundamental domain along the line connecting the upper
two fixed points and gluing the adjacent edges together, one arrives at a
ravioli- (cf.\ \cite{Quevedo:1996sv}) or pillow- (cf.\ \cite{Hebecker:2003jt})
like object which is flat everywhere except for the corners, i.e.\ the
orbifold fixed points, at which curvature and gauge field strength terms are
localized.

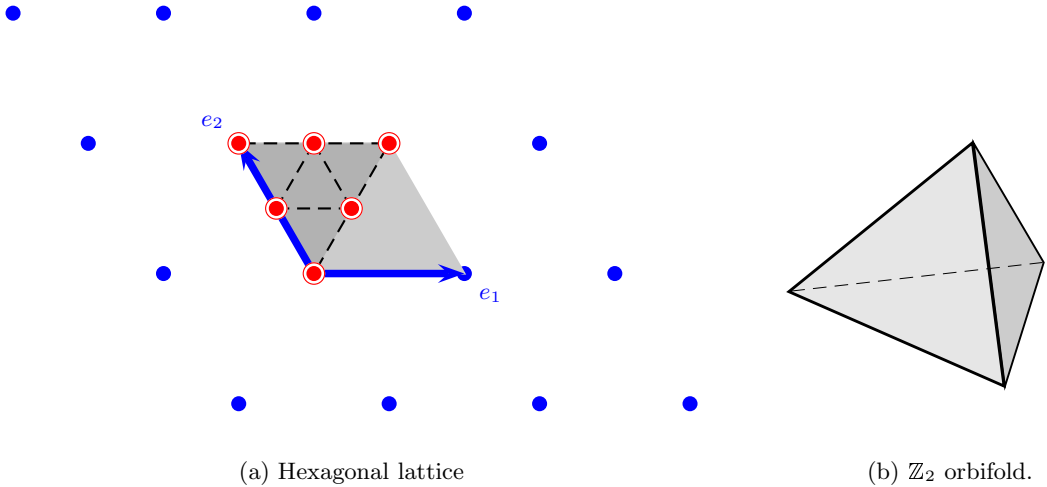
\begin{figure}[h]
\centerline{
\subfigure[Hexagonal lattice]{\psset{unit=2cm}
\begin{pspicture}(-2.1,-1)(2.6,1.8)
	\psdot*[dotstyle=*,dotsize=2mm,linecolor=blue](-0.5,-0.866025)
	\psdot*[dotstyle=*,dotsize=2mm,linecolor=blue](-1,0)
	\psdot*[dotstyle=*,dotsize=2mm,linecolor=blue](-1.5,0.866025)
	\psdot*[dotstyle=*,dotsize=2mm,linecolor=blue](-2.,1.73205)
	\psdot*[dotstyle=*,dotsize=2mm,linecolor=blue](0.5,-0.866025)
	\psdot*[dotstyle=*,dotsize=2mm,linecolor=blue](0,0)
	\psdot*[dotstyle=*,dotsize=2mm,linecolor=blue](-0.5,0.866025)
	\psdot*[dotstyle=*,dotsize=2mm,linecolor=blue](-1.,1.73205)
	\psdot*[dotstyle=*,dotsize=2mm,linecolor=blue](1.5,-0.866025)
	\psdot*[dotstyle=*,dotsize=2mm,linecolor=blue](1,0)
	\psdot*[dotstyle=*,dotsize=2mm,linecolor=blue](0.5,0.866025)
	\psdot*[dotstyle=*,dotsize=2mm,linecolor=blue](0.,1.73205)
	\psdot*[dotstyle=*,dotsize=2mm,linecolor=blue](2.5,-0.866025)
	\psdot*[dotstyle=*,dotsize=2mm,linecolor=blue](2,0)
	\psdot*[dotstyle=*,dotsize=2mm,linecolor=blue](1.5,0.866025)
	\psdot*[dotstyle=*,dotsize=2mm,linecolor=blue](1.,1.73205)
\pspolygon[fillstyle=solid,linecolor=gray80,fillcolor=gray80](0,0)(1,0)(0.5,0.866025)(-0.5,0.866025)(0,0)
\pspolygon[fillstyle=solid,linecolor=black,fillcolor=gray70,linestyle=dashed](0,0)(0.5,0.866025)(-0.5,0.866025)(0,0)
\psline[linewidth=1mm,linecolor=blue]{->}(0,0)(1,0)
\psline[linewidth=1mm,linecolor=blue]{->}(0,0)(-0.5,0.866025)
\rput[tl](1.1,-0.1){{\blue $e_1$}}
\rput[br](-0.6,0.966){{\blue $e_2$}}
\psline[linecolor=black,linestyle=dashed](-0.25,0.433)(0.25,0.433)
\psline[linecolor=black,linestyle=dashed](-0.25,0.433)(0,0.866025)
\psline[linecolor=black,linestyle=dashed](0.25,0.433)(0,0.866025)
	\psdot*[dotstyle=o,dotsize=3mm,linecolor=red](0,0)
	\psdot*[dotstyle=o,dotsize=3mm,linecolor=red](-0.5,0.866025)
	\psdot*[dotstyle=o,dotsize=3mm,linecolor=red](0.5,0.866025)
	\psdot*[dotstyle=o,dotsize=3mm,linecolor=red](0,0.866025)
	\psdot*[dotstyle=o,dotsize=3mm,linecolor=red](-0.25,0.433)
	\psdot*[dotstyle=o,dotsize=3mm,linecolor=red](0.25,0.433)
	\psdot*[dotstyle=*,dotsize=2mm,linecolor=red](0,0)
	\psdot*[dotstyle=*,dotsize=2mm,linecolor=red](-0.5,0.866025)
	\psdot*[dotstyle=*,dotsize=2mm,linecolor=red](0.5,0.866025)
	\psdot*[dotstyle=*,dotsize=2mm,linecolor=red](0,0.866025)
	\psdot*[dotstyle=*,dotsize=2mm,linecolor=red](-0.25,0.433)
	\psdot*[dotstyle=*,dotsize=2mm,linecolor=red](0.25,0.433)
\end{pspicture}\psset{unit=1cm}
}
\quad
\subfigure[$\Z2$ orbifold.]{\psset{unit=0.4cm}
\begin{pspicture}(-0,3.75)(14.14,14.14)
\psset{linestyle=solid,linewidth=0.03175,linecolor=white}
\psset{linewidth=0.0635,linecolor=black,fillstyle=solid,fillcolor=gray80}
\pspolygon(8.89,5.00)(10.18,9.12)(7.85,13.09)(8.89,5.00)
\psset{linewidth=0.09525,fillstyle=solid,fillcolor=gray90}
\pspolygon(1.71,8.15)(7.81,13.10)(8.86,5.01)(1.71,8.15)
\psset{linestyle=dashed,linewidth=0.03175,fillstyle=none}
\psline(1.79,8.17)(10.14,9.12)
\end{pspicture}\psset{unit=1cm}
}
}
\caption{The \Z2 orbifold of a torus based on a hexagonal lattice can be
envisaged as a tetrahedron.}
\label{fig:Tetrahedon}
\end{figure}

A special case of a 2D \Z2 orbifold arises if the lattice is hexagonal. Then the
edges of the fundamental cell can be glued together to a tetrahedron
\cite{Dixon:1986qv}. Again, the corners correspond to the fixed points (figure
\ref{fig:Tetrahedon}).

\subsection{Gauge group topography}

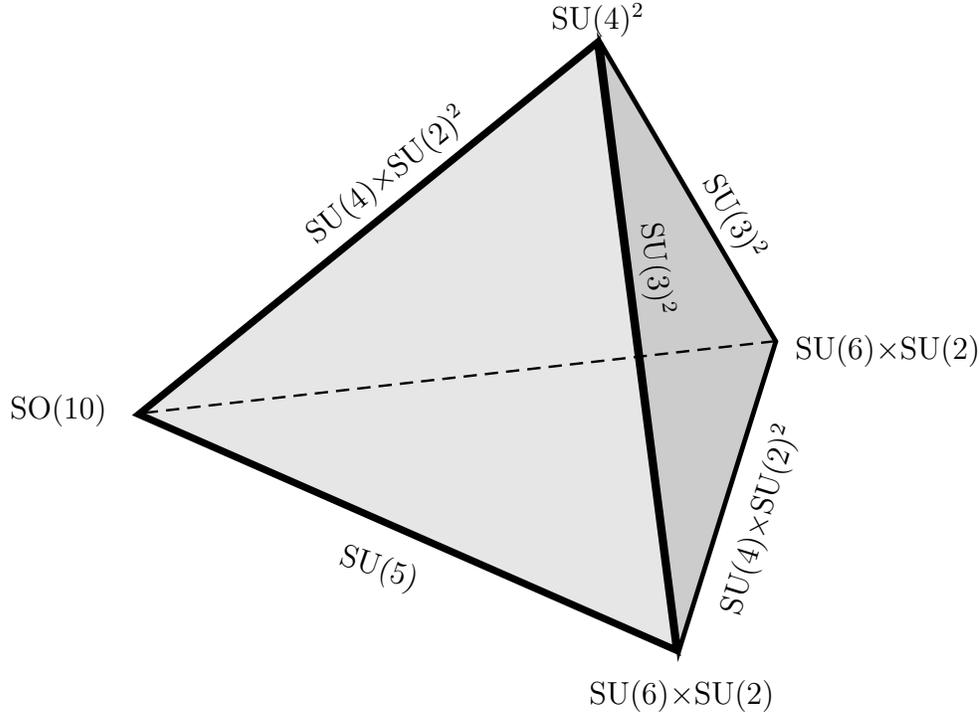
\begin{figure}[h]
\centerline{
\begin{pspicture}(-0cm,3.75cm)(14.14cm,14.14cm)
\psset{unit=1cm}
\psset{linestyle=solid,linewidth=0.03175,linecolor=white}
\psset{linewidth=0.0635,linecolor=black,fillstyle=solid,fillcolor=gray80}
\pspolygon(8.89,5.00)(10.18,9.12)(7.85,13.09)(8.89,5.00)
\rput[l](7.70,4.4){\large SU(6)$\times$SU(2)}
\rput[l](10.44,9.00){\large  SU(6)$\times$SU(2)}
\rput[l]{300}(9.3,11.3){\large  SU(3)$^2$}
\rput[l]{341}(4.40,6.27){\large SU(5)}
\rput[l]{39}(4,10.5){\large SU(4)$\times$SU(2)$^2$}
\rput[l](0.00,8.2){\large SO(10)}
\rput[l](7.2,13.4){\large SU(4)$^2$}
\rput[l]{73}(9.59,5.5){\large SU(4)$\times$SU(2)$^2$}
\rput[l]{6}(5.45,8.73){\large SU(5)}
\rput[l]{277}(8.5,10.70){\large SU(3)$^2$}
\psset{linewidth=0.09525,fillstyle=solid,fillcolor=gray90}
\pspolygon(1.71,8.15)(7.81,13.10)(8.86,5.01)(1.71,8.15)
\psset{linestyle=dashed,linewidth=0.03175,fillstyle=none}
\psline(1.79,8.17)(10.14,9.12)
\end{pspicture}
}
\caption{Gauge group topography. At different fixed points (corners of the
tetrahedron), \E8 gets broken to different subgroups (\U1 factors are
suppressed). At the edges we display the intersection of the two local gauge
groups realized at the corners.}
\label{fig:2D-Z2Example}
\end{figure}

Gauge theories in higher dimensions can exhibit a feature which we would like to
refer to as `gauge group topography'. That is, different gauge groups can be
realized at different points in internal space. The prime example is orbifolds
with non-trivial gauge embedding and discrete Wilson lines \cite{Ibanez:1986tp}.
Here, the 10D gauge group $\E8\times\E8$ gets broken to different subgroups at
different orbifold fixed points. Figure~\ref{fig:2D-Z2Example} illustrates  this
situation in a two-dimensional \Z2 orbifold. In non-prime orbifolds the
situation gets slightly richer in that there appear fixed planes which are
endowed with different gauge groups. This has lead to the notion `heterotic
brane world' \cite{Forste:2004ie}.

\begin{figure}[h]
\centerline{
\CenterEps{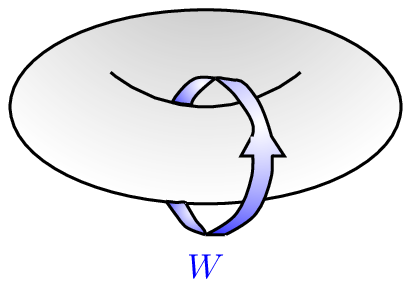}
\qquad\CenterObject{\begin{pspicture}(0,0)(1,2)
\psline(-0.4,1.5)(1,1.5)(0.9,1.4)
\psline(1,1.5)(0.9,1.6)
\rput[c](0.3,1.9){\large $\Z2$}
\end{pspicture}}
\qquad
\CenterEps{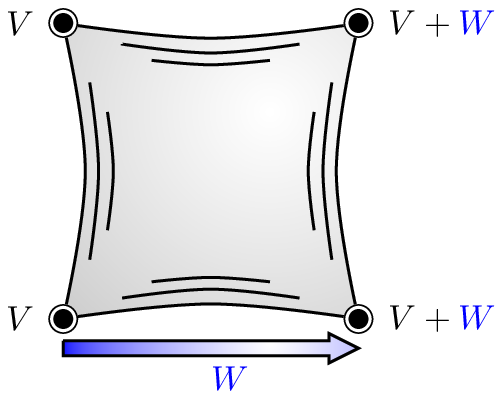}
}
\caption{A \Z2 orbifold in two steps. One compactifies first on a torus with
a discrete Wilson line $W$. Then, modding out a \Z2 symmetry of this torus leads
to an orbifold in which the local shifts differ by $W$.}
\label{fig:WilsonLine}
\end{figure}

In this review we refrain from giving a detailed, technical description on how
this works. All we need for the subsequent discussion is to understand how the
local gauge groups emerge. They are comprised out of the gauge bosons
corresponding to generators that fulfill certain (local) invariance conditions,
which can be recast schematically as
\begin{equation}\label{eq:invariance}
 \text{generator}\cdot \text{local shift}~=~0\mod 1\;.
\end{equation} 
The local shift \cite{GrootNibbelink:2003rc,Buchmuller:2006ik} introduced here
depends on the fixed point \cite{Ibanez:1986tp}. Since the invariance conditions
can be different at different fixed points, different gauge groups can live at
the various fixed points. For example, at the fixed points of the 
first twisted sector $T_1$, one finds
\begin{equation}
 V_\mathrm{local}~=~V+\text{Wilson lines}
 \qquad\leftrightarrow\qquad
 G_\mathrm{local}
 \;,
\end{equation}  
where the `Wilson lines' term depends on the specific fixed point.
This is illustrated in figure~\ref{fig:WilsonLine}.
In the field-theoretic description, one would say that the local gauge groups
are made up from the generators for which the corresponding gauge boson has a
non-vanishing profile at the fixed point. This also means that the coset,
$(\E8\times\E8)/G_\mathrm{local}$ with $G_\mathrm{local}$ denoting the local
gauge group, comprises generators where the gauge bosons' profile has to vanish
at the fixed point. From this point of view it is also clear what the massless
gauge interactions in 4D are: they are mediated by the bosons with flat profile,
i.e.\ which fulfill the invariance conditions \eqref{eq:invariance} everywhere.
In other words, the 4D gauge group emerges as the intersection of local gauge
groups in $\E8\times\E8$.

A simple way to think of this is by noting that gauge bosons are bulk fields,
and therefore feel what is going on at all positions in internal space. This is different
for localized fields, i.e.\ twisted matter, as we will discuss shortly in \ref{sec:LGU}.

As already said, we aim at deriving the standard model from heterotic orbifolds.
Since the 4D gauge group is the intersection of the local gauge groups at the
fixed points, this implies that, when the gauge group topography is non-trivial,
the local gauge groups at the fixed points have to be \emph{larger} than the
standard model, $G_\mathrm{local}\supset G_\mathrm{SM}$. This then leads to the
picture of `local grand unification' \cite{Buchmuller:2005sh}.

\subsection{Local grand unification}
\label{sec:LGU}

From the requirement $G_\mathrm{local}\supset G_\mathrm{SM}$ we see immediately that the
local gauge groups $G_\mathrm{local}$ can be groups that have been discussed in
the context of grand unification, such as the Pati-Salam group
$G_\mathrm{PS}=\SU4\times\SU2_\mathrm{L}\times\SU2_\mathrm{R}$, \SU5, and
\SO{10}. Recalling that fields confined to a region with
certain gauge symmetry have to furnish complete representations of this
symmetry, we see that matter fields localized in a region (or at a fixed point)
with GUT symmetry appear as GUT multiplets in the low-energy theory
\emph{although the gauge symmetry of the low-energy theory will be smaller}. 
Roughly speaking, brane fields only feel what's going on at where
they live and don't care about what's going on elsewhere. This statement
applies, in particular, to matter living in regions with \SO{10} symmetry: it
will furnish complete \SO{10} representations, i.e.\ a $\boldsymbol{16}$-plet
will give rise to a complete family of quarks and leptons. In other words, 
incomplete families cannot appear at these points nor do parts of these
families get projected out.

On the other hand, matter fields can also come from the bulk, i.e.\ the untwisted
sector. Such fields do, like the gauge bosons, feel the symmetry breaking at
every fixed point, and thus appear in split multiplets of the various 
local gauge groups. Together with what we
have discussed in the previous paragraph, this provides a very simple scheme
allowing to understand the simultaneous existence of complete and split
GUT multiplets in Nature. 

Based on these observations, we will below develop a strategy to construct
particle physics models in string theory, i.e.\ models with three chiral
generations plus one pair of Higgs below the compactification scale. The most
obvious possibility is to arrange things in such a way that there are three
$\boldsymbol{16}$-plets living in regions with \SO{10} symmetry plus a pair of
Higgs multiplets coming from the bulk. That is, we are going to look at models
in which there are orbifold fixed points at which $\E8$ gets broken to \SO{10}
and which give rise to localized $\boldsymbol{16}$-plets. The simplest (and
therefore best studied) setting, the \Z3 orbifold, fails to satisfy this
criterion (cf.\ \cite{Giedt:2000bi}). As we shall see later in
section~\ref{sec:ApproachingMSSM}, the situation improves when one turns to more
complicated settings, such as the \Z6-II orbifold. Before presenting this, we
need to discuss certain aspects of the structure of orbifold vacua.

\subsection{Orbifold vacua}
\label{sec:OrbifoldVacua}

As we have seen, an orbifold is defined by its geometry and its gauge embedding.
That is, given the geometry, shift and Wilson lines, the orbifold model is, in a
certain way, fixed: the massless spectrum as well as the couplings between the
various states are determined and calculable. However, as we shall explain now,
this does not fix the phenomenological properties of the model completely.
Rather, a given orbifold admits many vacua. The orbifold point, i.e.\ the point
in field space where all massless charged fields have zero vacuum expectation
values turns out not to correspond to a vacuum in most of the models with
non-standard embedding. Generically one of the \U1 factors appears anomalous.
This means that, at one-loop, a Fayet-Iliopoulos (FI) $D$-term is induced
\cite{Dine:1987xk}. Hence the orbifold point can be thought of as a saddle 
point in field space. To cancel the FI term, certain fields with charge opposite
to the trace of the `anomalous' \U1 have to attain vacuum expectation values. In
supersymmetric vacua, these vacuum expectation values need to be consistent with
zero $F$- and (other) $D$-terms. The vacuum space of orbifolds can be analyzed
field-theoretically. It is well known that a supersymmetric gauge theory with
generic superpotential has supersymmetric vacua with $F_i=D_a=0$
\cite{Wess:1992cp} (otherwise it would not be so hard to build field-theoretic
examples which spontaneously break supersymmetry). To see this, recall that
solutions to the $F$ equations still admit complexified gauge transformations,
allowing to simultaneously achieve $D_a=0$ \cite{Ovrut:1981wa}.\footnote{In
principle, these arguments tell us nothing about the expectation value of the
superpotential, $\langle\mathscr{W}\rangle$. For Calabi-Yau manifolds, which are
designed to give supersymmetric Minkowski vacua, one would expect vanishing $F$-
and $D$-terms and zero $\mathscr{W}$ simultaneously. Therefore one might
speculate that, since orbifolds are little cousins of Calabi-Yau manifolds, 
$\langle\mathscr{W}\rangle$ will vanish at the perturbative level if the $F$ and
$D$-equations are satisfied. As we shall discuss later in
section~\ref{sec:MiniLandscape}, we have verified explicitly that this is so, at
least to a certain order in couplings, which we restricted ourselves to for
technical reasons. }

What can we say about the moduli space of an orbifold model? The answer to this 
question is not straightforward because there are different branches of
vacua. There are supersymmetric vacua with a moduli space of large
dimensionality, i.e.\ with many flat directions. For instance, it was found that in a 
specific \Z6-II model the SM singlets from the sectors $U_1$, $U_2$,
$T_2$ and $T_4$  form exact (i.e.\ to all orders) $F$- and $D$-flat directions
\cite{Buchmuller:2006ik}, provided that the VEVs of the SM singlets from
the other sectors vanish. In this case, the fields fulfill what is sometimes
called the `stringent' $F$-flatness conditions (cf.\ e.g.\
\cite{Giedt:2005vx}): the $F$-terms satisfy
\begin{equation}
 F_i
 ~\sim~
 \frac{\partial \mathscr{W}}{\partial\phi_i}
  ~\propto~ \langle\text{field from }T_1\text{ or }T_3\rangle~=~0\;,
\end{equation}
i.e.\ the $F_i$ (as well as the superpotential $\mathscr{W}$) vanish to all orders
independently of the $U_1$, $U_2$, $T_2$ and $T_4$ VEVs.  At a generic point in
such vacuum, all gauge factors will be broken; the dimension of moduli space is
$\mathcal{O}(10-100)$.
It is easy to check that such vacua exist in many \Z6-II models. 
However, once a field from $U_3$, $T_1$
or $T_3$ attains a VEV, the situation changes: the above mentioned SM singlets cease
to form flat directions. The dimension of the moduli space is then much smaller,
in particular, it can be 0-dimensional, i.e.\ there can be isolated
supersymmetric vacua in field space. (The stabilization of the geometric moduli
and the dilaton will be briefly discussed below.) To find such vacua, one has to
find isolated solutions of the $F$ equations, and then use the freedom of
complexified gauge transformations to achieve $D_a=0$ (cf.\ the discussion in 
section~6 of~\cite{Buchmuller:2006ik}).

\begin{figure}[h]
\begin{center}
\begin{pspicture}(-6,-3)(6,3)
\rput[c](0,2){\rnode[c]{op}{\psframebox{\begin{tabular}{c}
orbifold point\\ (saddle point)
\end{tabular}}}}
\rput[c](-5,-2){\rnode[c]{v1}{\psframebox{\begin{tabular}{c}
supersymmetric\\ vacuum with
\\ dimension $\mathcal{O}(100)$
\end{tabular}}}}
\rput[c](-1,-2){\rnode[c]{v2}{\psframebox{\begin{tabular}{c}
another\\
supersymmetric\\ vacuum 
\end{tabular}}}}
\rput[c](2,-2){\Huge{\textbf{\dots}}}
\rput[c](5,-2){\rnode[c]{v3}{\psframebox{\begin{tabular}{c}
supersymmetric\\ vacuum with
\\ dimension $0$
\end{tabular}}}}
\ncline{->}{op}{v1}
\ncline{->}{op}{v2}
\ncline{->}{op}{v3}
\end{pspicture}
\end{center}
\caption{Different types of supersymmetric  vacua of a given orbifold model.}
\label{fig:SupersymmetricOrbifoldVacua}
\end{figure}
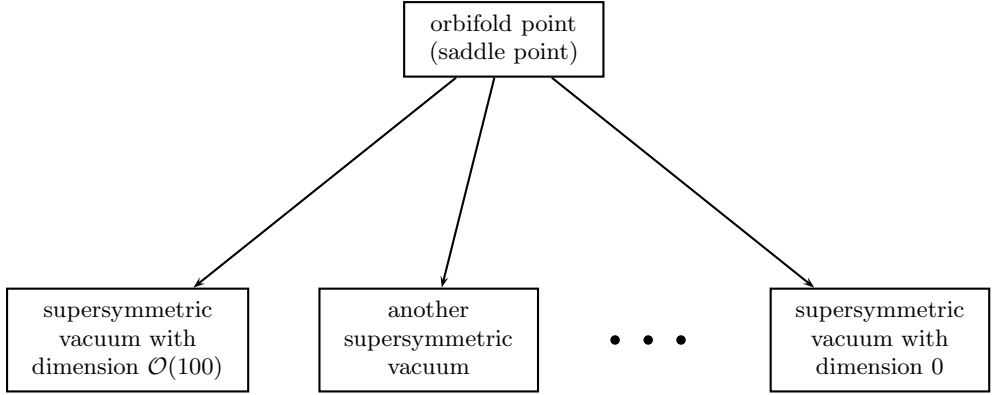

In conclusion, a given orbifold model has several branches of vacua
with different properties (figure \ref{fig:SupersymmetricOrbifoldVacua}). The
orbifold point is a point in field space around which the theory can be
expanded, but does in general not correspond to a vacuum configuration. As long
as the field VEVs are not too large, one might arguably retain control over the
theory. Potential obstacles have been identified (e.g.\ \cite{Cvetic:1998gv}),
and more work in this direction will be needed. But it is also clear that
certain features, such as the chiral spectrum, will not be affected by moving in
moduli space.

Before turning to the search for realistic vacua in orbifold models, let us
remark that to figure out how many vacua of a given dimension exist in an
orbifold model would represent an important piece of information in the context
of the landscape discussion \cite{Susskind:2003kw}. We will not attempt to
perform this analysis here, but for the subsequent discussion it is important to
keep in mind that counting orbifold models should not be confused with counting
string vacua; the actual number of orbifold vacua is certainly larger than the
number of models by many orders of magnitude. We will also discuss later in some
detail (section~\ref{sec:BlowUp}) that one cannot strictly distinguish between orbifold vacua and
Calabi-Yau vacua: giving vacuum expectation values to certain (twisted) fields
corresponds to blowing up the orbifold, i.e.\ smoothening out the singularities
associated to the orbifold fixed points.

\subsection{Properties of the effective field theories derived from orbifolds}
\label{sec:CouplingsAndCo}

 Given the field content of the orbifold, the next step is to study 
interactions of the theory. Couplings on orbifolds are governed by certain
selection rules \cite{Hamidi:1986vh,Dixon:1986qv}. From the effective field
theory perspective, these rules are
\begin{itemize}
 \item gauge invariance,
 \item (non-Abelian) discrete symmetries, and
 \item discrete $R$-symmetries.
\end{itemize}
The selection rules for the \Z6-II orbifold are summarized in
\cite{Buchmuller:2006ik} (a careful discussion of the space-group rule
is presented in \cite{Lebedev:2007hv}). They govern the couplings at the
orbifold point.

If one, as discussed above, moves away from the orbifold point in moduli space,
one obtains further, effective couplings, induced by the fields that attain
VEVs, which we call $s_i$ in what follows. The emerging picture is very similar
to the Froggatt-Nielsen scheme \cite{Froggatt:1978nt}, i.e.\ one has different
couplings arising at different orders in the $s_i$ fields. As long as one stays
close to the orbifold point, the VEVs $\langle s_i\rangle$ are small in string
units, and the effective couplings exhibit a hierarchical structure. This is
perhaps the most plausible way of generating hierarchical (Yukawa) couplings:
there are strong arguments by `t Hooft that hierarchically small parameters $y$
are only natural if in the limit where $y\to0$ a new symmetry appears
\cite{'tHooft:1979bh}. Orbifolds have this property: the orbifold point has many
symmetries some of which get broken by the $\langle s_i\rangle$ VEVs, and get
restored as $\langle s_i\rangle\to 0$. The fact that there is a huge literature
on rather compelling models in which realistic Yukawa couplings are explained in
this way (cf.\ e.g.\ \cite{Leurer:1993gy,Binetruy:1996xk}) gives further credit
to this picture. A specific example of an orbifold that leads to hierarchical
couplings will be discussed in section \ref{sec:BenchmarkModel}. In conclusion,
from the fact that the Yukawa couplings observed in Nature exhibit (strong)
hierarchies one might conclude that we live somewhere not too far from an
orbifold point in field space.

The main lesson is that, once the geometry and gauge embedding are fixed, the
model \emph{at the orbifold point} is completely determined. In particular, 
unlike in the field-theoretic constructions, the spectrum is fixed by the
compactification
and one cannot `put by hand'
representations and/or couplings. For configurations close to the orbifold
point, one can use perturbation theory in order to calculate the effective
couplings.

\section{Approaching the MSSM}
\label{sec:ApproachingMSSM}

In the previous sections we have seen that orbifold compactifications provide us
with everything needed for deriving the standard model from string theory: gauge
groups smaller than $\E8\times\E8$ and chiral matter. We have also explained
that, by exploiting the gauge group topography of orbifolds, one can set up
models of `local grand unification' which can explain the appearance of larger
representations while the gauge group is broken to smaller pieces. A crucial
ingredient in this scheme is $\boldsymbol{16}$-plets localized at fixed points
with \SO{10} symmetry.

In the \Z6-II orbifold, there are two (local) shifts that produce a local
\SO{10} GUTs with local $\boldsymbol{16}$-plets (cf.\ \cite{Katsuki:1989cs}),
\begin{eqnarray}
 V^{ \SO{10},1} &= &
 \left(\tfrac{1}{3},\,\tfrac{1}{2},\,\tfrac{1}{2},\,0,\,0,\,0,\,0,\,0\right)\,\left(\tfrac{1}{3},\,0,\,0,\,0,\,0,\,0,\,0,\,0\right)
 \;,
 \nonumber \\
 V^{\SO{10},2}& = &
 \left(\tfrac{1}{3},\,\tfrac{1}{3},\,\tfrac{1}{3},\,0,\,0,\,0,\,0,\,0\right)\,\left(\tfrac{1}{6},\,\tfrac{1}{6},\,0,\,0,\,0,\,0,\,0,\,0\right)
 \;. \label{eq:so10shifts}
\end{eqnarray}
Our strategy will be to construct models based on the \SO{10} shifts and see to
which extent they can be consistent with particle physics. More precisely, we
will conduct a search for models with vacua which can give rise to the  MSSM at
energies below $M_\mathrm{GUT}$ within this set of orbifolds.

\subsection{String derived orbifold GUTs}

The shift $V^{ \SO{10},2}$ has been used in the first string-derived orbifold
GUT models~\cite{Kobayashi:2004ud,Kobayashi:2004ya}. There, various models with
Pati-Salam symmetry $G_\mathrm{PS}$ at the orbifold point were constructed.
These models exhibit three chiral generations as well as the Higgs fields,
required to break $G_\mathrm{PS}\to G_\mathrm{SM}$, and the electroweak Higgs
doublets. There are also further interesting features such as order one Yukawa
couplings for the third generation and a $D_4$ flavor symmetry for the two light
generations. It has also been attempted to use the shift $V^{ \SO{10},2}$ in a
model with standard model gauge symmetry where three generations arise from
three $\boldsymbol{16}$-plets localized at \SO{10} fixed points
\cite{Buchmuller:2004hv}. However, all of the the models of
\cite{Kobayashi:2004ud,Kobayashi:2004ya,Buchmuller:2004hv} have a common
problem: the appearance of exotic states at low energies. 

In fact, we have seen that, in order to get \E8 broken to the standard model, we
need fixed points with a symmetry that is not $\SO{10}\supset G_\mathrm{SM}$.
These fixed points will carry twisted states, giving rise to non-SM matter. The
only situation in which these exotics can be harmless is when they are vector-like
w.r.t.\ $G_\mathrm{SM}$: in this case, certain SM singlet fields might attain
VEVs which, provided that suitable couplings exist, lead to mass terms for the
exotics. To show that the required couplings and singlets exist is of course not
sufficient, one must also verify that the desired singlet VEVs lie at (local)
minima of the potential. In the models \cite{Kobayashi:2004ud,Kobayashi:2004ya}
it was found that the required couplings do not exist.\footnote{By now it has
become clear that, with the correct selection rules \cite{Lebedev:2007hv}, the
couplings do exist. It might therefore be worthwhile to revisit the models.} In
the simple models where all three generations originate from
$\boldsymbol{16}$-plets localized at \SO{10} fixed points, in particular in the
one presented in \cite{Buchmuller:2004hv}, it is found that, if one insists on
hypercharge being normalized as in GUTs, there are always chiral exotics
\cite{Buchmuller:2006ik,Lebedev:2006kn}. This statement applies to $\Z{N\le8}$
orbifolds. One is hence lead to consider schemes in which at least one family
does not come from a localized $\boldsymbol{16}$-plet.

In summary, the first string derived orbifold GUTs have many interesting and
appealing features. The main drawback is the appearance of unwanted states.
We would like to stress that constraints from the model to be global
(rather than local) has important consequences for the phenomenological
viability of string compactifications. If we were to neglect the states from
certain fixed points, it would be easy to obtain constructions without unwanted
exotics. 

Recently there has been progress in F-theory compactifications of string theory
\cite{Donagi:2008ca,Beasley:2008dc}. As of now, only local models have been
discussed. It remains to be seen if this approach allows to build globally
consistent and simultaneously phenomenologically viable models. This is
an interesting question because these constructions are related to D-brane
models where getting the exact MSSM spectrum appears to be notoriously
problematic. 
On the other hand, some constructions in this scheme
\cite{Donagi:2008ca} are dual to heterotic compactifications, where models with
the exact MSSM spectrum can indeed be obtained, as we shall now discuss.

\subsection{MSSM from the heterotic string}
\label{sec:MSSMfromString}

There are \Z6-II models which can indeed exhibit the exact MSSM spectrum below
$M_\mathrm{GUT}$; an example has been presented
in \cite{Buchmuller:2005jr,Buchmuller:2006ik}. It is based on $V^{\SO{10},1}$;
the Wilson lines are chosen such that 
\begin{equation}
 \text{gauge group}~=~(G_\mathrm{SM}\subset\SO{10})\times\text{extra factors}
\end{equation}
and
\begin{equation}
 \text{spectrum}~=~3\times\text{generation}+\text{vector-like w.r.t.\
 $G_\mathrm{SM}$}
\end{equation}
at the orbifold point. It has been demonstrated that the exotic states can be
decoupled: (i) there exist couplings of the form
\begin{equation}\label{eq:exotics1}
 x_j\,\bar x_k\,\prod\limits_{i} s_i\;,
\end{equation}
where $x_j$ and $\bar x_k$ denote the vector-like exotics and $s_i$ are standard
model singlets. (ii) It can be shown that the $s_i$ VEVs are consistent with
supersymmetry (as discussed in section~\ref{sec:OrbifoldVacua}, one essentially
has to check that the $s_i$ VEVs are consistent with vanishing $D$-terms).

In this model, two generations stem from localized $\boldsymbol{16}$-plets while
the third generation comes from states from the untwisted or higher twisted
sectors, i.e.\ fields that live in the bulk or at orbifold fixed planes. 

This model has been analyzed in detail in \cite{Buchmuller:2006ik} and in
\cite{Buchmuller:2007qf}, where the advantages of the local GUT  representations
have been exploited and the (higher-dimensional) anomaly constraints have been
checked. We will refrain from reviewing this discussion here in detail since
below we will discuss a very similar model that shares many important properties
with the one considered here.

\subsection{The heterotic Mini-Landscape}
\label{sec:MiniLandscape}

The scheme of local grand unification can be utilized in the search for models
with realistic features. Given the model discussed above
(section~\ref{sec:MSSMfromString}), one might wonder whether there are more 
models with similar properties.
In what follows, we discuss the results of a scan over
\Z6-II orbifolds with local \SO{10} GUT structures.
For each of the \SO{10} shifts \eqref{eq:so10shifts}, the following steps were
performed:
\begin{dingautolist}{"0C0} 
 \item \label{step1}
 Generate Wilson lines $W_3$ and $W_2$.\\
 One of the \SO{10} shifts \eqref{eq:so10shifts} is chosen and all two Wilson
 line models are constructed using the methods described
 in~\cite{Giedt:2000bi} and in the appendix
 of~\cite{Nilles:2006np}. These models can be separated into two cases: either
 these two Wilson lines are of order two ($W_2$ and $W_2'$) or one Wilson line
 is of order three ($W_3$) and one of order two ($W_2$).\footnote{A comment for
 experts is in order: the \Z6-II
 orbifold does not allow for (generalized) discrete torsion
 \cite{Vafa:1986wx,Gaberdiel:2004vx} (cf.\ \cite{Ploger:2007iq}). 
 Hence the ansatz of~\cite{Giedt:2000bi,Nilles:2006np} is
 sufficient to construct all different models. It is, however, clear that, once
 one changes the geometry, for instance goes to non-factorizable lattices
 \cite{Faraggi:2006bs,Forste:2006wq,Takahashi:2007qc} new models will arise.}
 \item \label{step2}
  Identify ``inequivalent'' models.\\
  As is well-known, different looking Wilson lines can still lead to equivalent
  models because they are related by Weyl reflections and/or lattice
  translations. However, as the Weyl group is huge, it would be extremely time
  consuming to probe whether two Wilson lines are equivalent or not. In our
  computations, we therefore adopt a rather pragmatic notion of equivalence: two
  models are taken to be ``equivalent'' if their spectra coincide.\footnote{ In 
  practice, we compare non-Abelian representations and the number of
  non-Abelian  singlets.} This underestimates the true number of models.
 \item \label{step3}
  Select models with $G_\mathrm{SM} \subset \SU5 \subset \SO{10}$.\\
  We select models with $\SU3\times\SU2$ that fit into the local \SO{10}.
  There will always be an $\SU5$ subgroup of \SO{10} that contains 
  $\SU3\times\SU2$. Since (Abelian) \Z{N} orbifolds do not reduce the rank, 
  all \U1 factors survive orbifolding, and hence the selected models will have 
  $G_\mathrm{SM} \subset \SU5 \subset \SO{10}$.
 \item \label{step4}
  Select models with three net $(\boldsymbol{3},\boldsymbol{2})$.\\
  Having identified $\SU3_\mathrm{C}\times\SU2_\mathrm{L}\subset G_\mathrm{SM}$
  in the previous step, we project now on models that have a net number of three
  $(\boldsymbol{3},\boldsymbol{2})$ (quark doublets).
  Here, a net number means that we also allow for situations where the spectrum
  contains four $(\boldsymbol{3},\boldsymbol{2})$  plus one
  $(\boldsymbol{\overline{3}},\boldsymbol{2})$, for example. At this level, 
  this amounts to requiring three families.
 \item 
  \label{step5} 
  Select models with non-anomalous $\U1_{Y} \subset \SU5$.\\
  This ensures that the hypercharge chosen previously in step~\ref{step3} is
  non-anomalous. Technically, this is achieved by demanding that the respective
  generators are orthogonal, i.e.\ $\mathsf{t}_Y \cdot \mathsf{t}_\mathrm{anom} =
  0$. A non-anomalous hypercharge is necessary, because an anomalous one would be
  broken at the high scale due to the presence of the FI $D$-term (cf.\
  section~\ref{sec:OrbifoldVacua}), 
  resulting in 
  electroweak symmetry breaking at a high energy scale.
 \item Select models with net 3 SM families + Higgses + vector-like.\\
  In the last step models are selected which have the chiral matter content
  of the MSSM, i.e.\ three generations of quarks and leptons.  Additionally, the
  models are allowed to have vector-like exotics. In order for some exotics to
  be vector-like they either have to form real representations or they have to
  come in pairs of some representations plus their complex conjugates. Then, it
  is in principle possible to write down a mass term for these exotics with a
  very high mass such that the exotics decouple from the low energy effective
  theory. Note, however, that the couplings in the superpotential relevant for
  the mass terms can not be put in by hand, but they have to be derived from
  string theory, as discussed in section~\ref{sec:CouplingsAndCo}.
\end{dingautolist}
It turns out that in these models almost $1\,\%$ has the MSSM spectrum plus
vector-like exotics (table \ref{tab:Statistics}).

\begin{table}[h!]
\centerline{
\begin{tabular}{|l||l|l|}
\hline
 criterion & $V^{\SO{10},1}$ & $V^{\SO{10},2}$ \\
\hline
&&\\[-0.3cm]
 \ding{"0C1}  inequivalent models with 2 Wilson lines
  &$22,000$ & $7,800$   \\[0.2cm]
  \ding{"0C2} SM gauge group $\subset$ SU(5) $\subset$ SO(10)
  &3,563 &1,163 \\[0.2cm]
  \ding{"0C3} 3 net $(\boldsymbol{3},\boldsymbol{2})$
  &1,170 &492 \\[0.2cm]
  \ding{"0C4} non-anomalous $\U1_{Y}\subset \SU5 $
  &528 &234 \\[0.2cm]
  \ding{"0C5}  spectrum $=$ 3 generations $+$ vector-like
  &128 &90 
  \\
\hline
\end{tabular}
}
\caption{Statistics of \Z6-II orbifolds based on the SO(10) shifts
\eqref{eq:so10shifts} with two Wilson lines. \label{tab:Statistics} }
\label{tab:statistics}
\end{table}

The main conclusion that one can draw from these statistics is that heterotic
orbifolds with local \SO{10} structures are a particularly ``fertile'' scheme
for producing models that are close to the MSSM. (The question ``how close?''
will be studied in the next subsection.) To see this, let us compare our
statistics with other MSSM searches in the
literature. In certain types of intersecting D-brane models, it was found that
the probability of obtaining the SM gauge group and three generations of quarks
and leptons, while allowing for chiral exotics, is less than $10^{-9}$
\cite{Gmeiner:2005vz,Douglas:2006xy}. The criterion which comes closest to the
requirements imposed in \cite{Gmeiner:2005vz,Douglas:2006xy} is \ding{"0C3}.  We
find that within our sample the corresponding probability is 6\,\%. In
\cite{Dijkstra:2004cc,Anastasopoulos:2006da}, orientifolds of Gepner models were
scanned for chiral MSSM matter spectra, and it was found that the fraction of
such models is $4 \times 10^{-14}$. These constructions contain the MSSM matter
spectrum plus, in general, vector-like exotics.   This is most similar to step
\ding{"0C5} in our analysis where we find 218 models out of a total of $3
\times 10^4$ or 0.7\,\%. 
Additionally, approximately 0.6\,\% of our models have
the MSSM spectrum at low energies with all vector-like exotics decoupling along
$D$-flat directions. Note also that, in all of our 218 models, hypercharge is
normalized as in standard GUTs and thus consistent with gauge coupling
unification. Let us also remark that all such models are very similar to the one
discussed in section~\ref{sec:MSSMfromString}; in particular, they exhibit a
$2+1$ family structure, i.e.\ there are two very similar families coming from
localized $\boldsymbol{16}$-plets living at \SO{10} fixed points plus one very
different family scattered over the bulk and fixed planes. Note also that the
identification of the third family is not straightforward as there is a mixing
with vector-like states carrying SM quantum numbers.

We would like to remark that our GUT-based strategy to determine the hypercharge is, of
course, not unique. One could instead express $\U1_Y$ as an arbitrary linear
combination of all \U1's (not only of those embedded in the local GUT symmetry),
such that it gives the correct values of hypercharge to the MSSM particles. This
approach was followed in \cite{Raby:2007yc}. The authors of \cite{Raby:2007yc}
find that the majority of the models at step~\ding{"0C3} allow for a definition
of a non-anomalous $\U1_Y$. However,  only in 12\,\% of those models, hypercharge
is in harmony with coupling unification. That means, in particular, that even in
a more general scheme, relaxing the demand $\U1_{Y}\subset \SU5$, (almost) only
those 223 models at step~\ding{"0C5} of our search meet all the phenomenological
properties we require.

Given these models, we study their properties. We analyze the question whether
the appearance of MSSM gauge group and spectrum is correlated to other
properties of the model. To be specific, we study the properties of the
so-called hidden sector. 

It is clear that the pure MSSM does not represent a complete setup; one has to
amend it by a sector that is responsible for supersymmetry breakdown, which is
usually called the `hidden sector' (cf.\ \cite{Nilles:1983ge}). This is because
settings in which supersymmetry is broken by auxiliary VEVs of MSSM fields are
challenged by observation, if not ruled out.  Arguably, the most appealing types of hidden sectors are those in
which the scale of supersymmetry breakdown is explained by dimensional
transmutation, as in the scheme of gaugino condensation
\cite{Nilles:1982ik,Ferrara:1982qs,Derendinger:1985kk,Dine:1985rz}. The 
gravitino mass, setting the scale of MSSM soft supersymmetry breaking terms, is
\begin{equation}\label{eq:m32}
 m_{3/2}~\sim~\frac{\Lambda^3}{M_\mathrm{Pl}^2} \;,
\end{equation}
where the gaugino condensation scale $\Lambda \equiv \langle \lambda \lambda
\rangle^{1/3} $ is given by the renormalization group (RG)  invariant scale of
the condensing gauge group,
\begin{equation}
 \Lambda~\sim~ 
 M_\mathrm{GUT}\,\exp \left(
 -\frac{1}{2\beta}\,\frac{1}{g^2(M_\mathrm{GUT})}
 \right) \;.
\label{eq:Lambda}
\end{equation}
Here $\beta$ denotes the $\beta$-function coefficient, which depends on the
gauge group and the matter content.

In string theory, the question of supersymmetry breakdown is usually closely
related to the mechanism of moduli stabilization. A condensing gauge group often
leads to the dilaton run-away problem \cite{Dine:1985kv}. This problem can be
avoided in various ways. Here we consider the scheme of K\"ahler stabilization
where the tree level K\"ahler potential of the dilaton gets amended by a
non--perturbative correction, $K=-\ln (S+ \bar S) + \Delta K_\mathrm{np}$. The
form of this correction has been studied in \cite{Binetruy:1996xj,Casas:1996zi,Lowen:2008fm}
(for a review see \cite{Gaillard:2007jr}).
With a favorable choice of the parameters, the dilaton can be stabilized at a
realistic value $\re S \approx 2$. In this review, we do not discuss how
plausibly the parameters will fall into favorable ranges;
we will just assume that there exists a mechanism that successfully stabilizes
the dilaton  while breaking supersymmetry,
\begin{equation}
 F_S~\sim~\frac{\Lambda^3}{M_\mathrm{Pl}} \;,
\end{equation}
which is in agreement with relation \eqref{eq:m32}. Note that in this scheme,
as a consequence of $T$-duality invariance of the non-perturbative
superpotential terms, the
geometric moduli, in particular the K\"ahler (or volume) moduli get fixed as
well \cite{Font:1990nt,Nilles:1990jv}.

\begin{figure}[!h!]
\centerline{\includegraphics{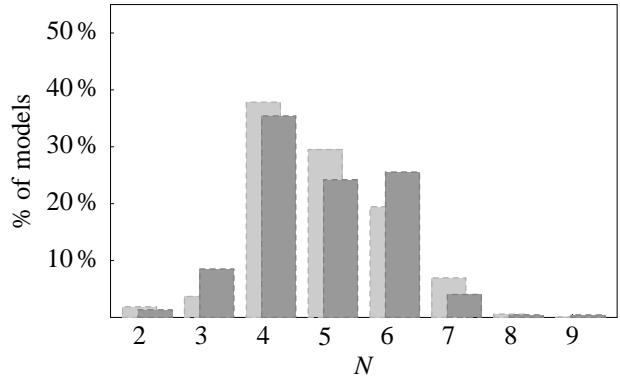}}
\caption{Number of models vs.\ 
the size of largest gauge group in the hidden sector. 
$N$ labels  $\SU{N}$, $\SO{2N}$, $\E{N}$ groups.
The background corresponds to step \ding{"0C1}, while the foreground
corresponds to step \ding{"0C5}.
}
\label{fig:histogram1}
\end{figure}

\begin{figure}[!h!]
\centerline{\includegraphics{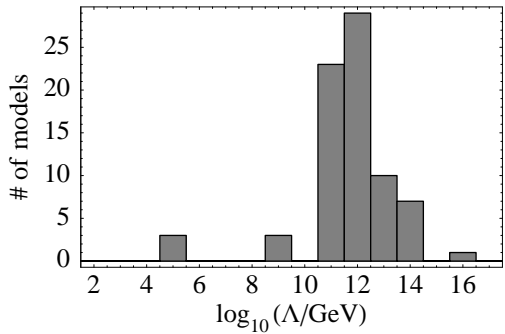}}
\caption{Number of models vs.\ (naive) scale of gaugino
condensation.}
 \label{fig:Lambda}
\end{figure}

In string models, of course, we cannot `invent' hidden sectors, but we have to
live with what strings give us. The non-Abelian subgroups of the second \E8
factors in the 218 models with chiral MSSM spectrum (last line
in table~\ref{tab:statistics}) were analyzed in~\cite{Lebedev:2006tr}. The result
is illustrated in figure~\ref{fig:histogram1}.
We see that there is a statistical preference for $\SU{N}$ and $\SO{2N}$ groups
with $N$ ranging between 4 and 6. By calculating the corresponding
$\beta$-functions, we obtain an estimate on the scale of $\Lambda$
\eqref{eq:Lambda} (figure~\ref{fig:Lambda}). 
This estimate is very rough, it neglects, in particular, string threshold
corrections \cite{Dixon:1990pc,Mayr:1993mq,Stieberger:1998yi}.\footnote{ 
Some progress in understanding  threshold corrections in the presence of Wilson
lines has been reported recently in \cite{Kim:2007jg}.}
It is, nevertheless, remarkable that the distribution shows a clear statistical
preference for intermediate scale supersymmetry breaking. This might be
interpreted as a top-down motivation for low-energy
supersymmetry~\cite{Lebedev:2006tr}.

\subsection{How good are the Mini-Landscape models?}
\label{sec:BenchmarkModel}

Having obtained about 200 models with exact MSSM spectra, we now discuss their
phenomenological properties. There are two kinds of questions that we will
address:
\begin{enumerate}
 \item Can we reproduce/accommodate all the features of the MSSM?
 \item Does string theory give us even more, i.e.\ are there stringy mechanisms
 at work that help to solve some of the MSSM puzzles (such as the $\mu$, strong
 CP and MSSM fine tuning problems)?
\end{enumerate}
In what follows we will answer the questions as far as possible. We will base
our discussion on a specific model -- the `benchmark' model 1A presented in
\cite{Lebedev:2007hv} -- yet the features discussed are shared with many models
of the Mini-Landscape.

As explained before, once the shift and Wilson lines are specified, the spectrum of
the model is fixed. The gauge group of the benchmark model is
\begin{equation}
 \text{gauge group @ orbifold point}~=~
 G_\mathrm{SM}\times\U1_\mathrm{B-L}\times[\SU4\times\SU2]
 \times\U1_\mathrm{anom}\times\U1^6\;.
\end{equation}
The quantum numbers of the massless states w.r.t.\ $\SU3\times\SU2\times[\SU4\times\SU2'] $ are shown
in table~\ref{tab:spectrum}.
\begin{table}[h]
\centerline{%
\begin{tabular}{|c|l|l|c|c|l|l|}
\hline
\# & irrep & label & & \# & irrep & label\\
\hline
 3 &
$\left(\boldsymbol{3},\boldsymbol{2};\boldsymbol{1},\boldsymbol{1}\right)_{(1/6,1/3)}$
 & $q_i$
 & &
 3 &
$\left(\overline{\boldsymbol{3}},\boldsymbol{1};\boldsymbol{1},\boldsymbol{1}\right)_{(-2/3,-1/3)}$
 & $\bar u_i$
 \\
 3 &
$\left(\boldsymbol{1},\boldsymbol{1};\boldsymbol{1},\boldsymbol{1}\right)_{(1,1)}$
 & $\bar e_i$
 & &
 8 &
$\left(\boldsymbol{1},\boldsymbol{2};\boldsymbol{1},\boldsymbol{1}\right)_{(0,*)}$
 & $m_i$
 \\
 4 &
$\left(\overline{\boldsymbol{3}},\boldsymbol{1};\boldsymbol{1},\boldsymbol{1}\right)_{(1/3,-1/3)}$
 & $\bar d_i$
 & &
 1 &
$\left(\boldsymbol{3},\boldsymbol{1};\boldsymbol{1},\boldsymbol{1}\right)_{(-1/3,1/3)}$
 & $d_i$
 \\
 4 &
$\left(\boldsymbol{1},\boldsymbol{2};\boldsymbol{1},\boldsymbol{1}\right)_{(-1/2,-1)}$
 & $\ell_i$
 & &
 1 &
$\left(\boldsymbol{1},\boldsymbol{2};\boldsymbol{1},\boldsymbol{1}\right)_{(1/2,1)}$
 & $\bar \ell_i$
 \\
 1 &
$\left(\boldsymbol{1},\boldsymbol{2};\boldsymbol{1},\boldsymbol{1}\right)_{(-1/2,0)}$
 & $h_d$
 & &
 1 &
$\left(\boldsymbol{1},\boldsymbol{2};\boldsymbol{1},\boldsymbol{1}\right)_{(1/2,0)}$
 & $h_u$
 \\
 6 &
$\left(\overline{\boldsymbol{3}},\boldsymbol{1};\boldsymbol{1},\boldsymbol{1}\right)_{(1/3,2/3)}$
 & $\bar\delta_i$
 & &
 6 &
$\left(\boldsymbol{3},\boldsymbol{1};\boldsymbol{1},\boldsymbol{1}\right)_{(-1/3,-2/3)}$
 & $\delta_i$
 \\
 14 &
$\left(\boldsymbol{1},\boldsymbol{1};\boldsymbol{1},\boldsymbol{1}\right)_{(1/2,*)}$
 & $s^+_i$
 & &
 14 &
$\left(\boldsymbol{1},\boldsymbol{1};\boldsymbol{1},\boldsymbol{1}\right)_{(-1/2,*)}$
 & $s^-_i$
 \\
 16 &
$\left(\boldsymbol{1},\boldsymbol{1};\boldsymbol{1},\boldsymbol{1}\right)_{(0,1)}$
 & $\bar n_i$
 & &
 13 &
$\left(\boldsymbol{1},\boldsymbol{1};\boldsymbol{1},\boldsymbol{1}\right)_{(0,-1)}$
 & $n_i$
 \\
 5 &
$\left(\boldsymbol{1},\boldsymbol{1};\boldsymbol{1},\boldsymbol{2}\right)_{(0,1)}$
 & $\bar \eta_i$
 & &
 5 &
$\left(\boldsymbol{1},\boldsymbol{1};\boldsymbol{1},\boldsymbol{2}\right)_{(0,-1)}$
 & $\eta_i$
 \\
 10 &
$\left(\boldsymbol{1},\boldsymbol{1};\boldsymbol{1},\boldsymbol{2}\right)_{(0,0)}$
 & $h_i$
 & &
 2 &
$\left(\boldsymbol{1},\boldsymbol{2};\boldsymbol{1},\boldsymbol{2}\right)_{(0,0)}$
 & $y_i$
 \\
 6 &
$\left(\boldsymbol{1},\boldsymbol{1};\boldsymbol{4},\boldsymbol{1}\right)_{(0,*)}$
 & $f_i$
 & &
 6 &
$\left(\boldsymbol{1},\boldsymbol{1};\overline{\boldsymbol{4}},\boldsymbol{1}\right)_{(0,*)}$
 & $\bar f_i$
 \\
 2 &
$\left(\boldsymbol{1},\boldsymbol{1};\boldsymbol{4},\boldsymbol{1}\right)_{(-1/2,-1)}$
 & $f_i^-$
 & &
 2 &
$\left(\boldsymbol{1},\boldsymbol{1};\overline{\boldsymbol{4}},\boldsymbol{1}\right)_{(1/2,1)}$
 & $\bar f_i^+$
 \\
 4 &
$\left(\boldsymbol{1},\boldsymbol{1};\boldsymbol{1},\boldsymbol{1}\right)_{(0,\pm2)}$
 & $\chi_i$
 & &
 32 &
$\left(\boldsymbol{1},\boldsymbol{1};\boldsymbol{1},\boldsymbol{1}\right)_{(0,0)}$
 & $s^0_i$
 \\
 2 &
$\left(\overline{\boldsymbol{3}},\boldsymbol{1};\boldsymbol{1},\boldsymbol{1}\right)_{(-1/6,2/3)}$
 & $\bar v_i$
 & &
 2 &
$\left(\boldsymbol{3},\boldsymbol{1};\boldsymbol{1},\boldsymbol{1}\right)_{(1/6,-2/3)}$
 & $v_i$
 \\
\hline
\end{tabular}
}
\caption{Spectrum. The quantum numbers under
$\SU3\times\SU2\times[\SU4\times\SU2']$ are shown in boldface; hypercharge and
\BmL\ charge appear as subscripts.  Note that the states $s_i^\pm$, $f_i$, $\bar
f_i$ and $m_i$ have different $B-L$ charges for different $i$, which we do not
explicitly list.}
\label{tab:spectrum}
\end{table}
Here we list also the $B-L$ (with $B$ and $L$ denoting baryon and lepton number,
respectively). This $B-L$ symmetry is related, but not identical, to the
canonical $B-L$ embedded in the local \SO{10}.
Indeed, the canonical $B-L$ symmetry turns out to be anomalous, i.e.\ the
corresponding generator $\mathsf{t}_{B-L}^{\SO{10}}$ is not orthogonal to the
generator of the anomalous \U1. It is, however, possible to define $\U1_{B-L}$
as a linear combination of $\mathsf{t}_{B-L}^{\SO{10}}$ with the other \U1
generators in such a way that
\begin{itemize}
 \item the charges of the members of the $\boldsymbol{16}$-plets are standard;
 \item the spectrum is three chiral generations plus vector-like w.r.t.\
 $G_\mathrm{SM}\times\U1_{B-L}$.
\end{itemize}
Given these properties, it is clear that, just like in usual GUTs, the $B-L$
symmetry can be used to distinguish Higgs from lepton doublets. The fact that we
had to redefine $B-L$ implies that the normalization is different from the one
in ordinary GUTs, which leads to the appearance of SM singlet fields with $B-L$
charge $q_{B-L}=\pm2$ (named $\chi$-plets in table~\ref{tab:spectrum}). This 
has important consequences, as we shall discuss next.

As already mentioned, one \U1 factor appears anomalous. As we
have discussed in section~\ref{sec:OrbifoldVacua}, this means that there is a
$D$-term that has to be cancelled. It has been checked that this can be achieved by
giving fields $\widetilde{s}_i$ a VEV which transform trivially under
$G_\mathrm{SM}\times\SU4$ and whose $B-L$ charge is either 0 or $\pm2$. As
explained in section~\ref{sec:OrbifoldVacua}, the $F$-term equations can be
solved simultaneously, hence we have obtained supersymmetric vacua. Because
$B-L$ is broken by two units, these vacua exhibit a matter parity symmetry
\cite{Dimopoulos:1981dw} $\mathbbm{Z}_2^\mathcal{M}$, which has the same
phenomenological implications as the usual MSSM $R$-parity: dangerous dimension four
proton decay operators do not exist and the lightest supersymmetric particle
(LSP), which is known to be an excellent dark matter candidate, is stable. 

Giving VEVs to the $\widetilde{s}$ fields has further crucial implications.
As we have seen, the spectrum (table~\ref{tab:spectrum}) after compactification,
i.e.\ at the orbifold point, contains, as desired, three generations of MSSM
matter as well as (unwanted) vector-like exotics.
It turns out that due to the $\langle\widetilde{s}_i\rangle$ VEVs the exotics
attain masses, i.e.\ there exist couplings of the form
\begin{equation}\label{eq:exotics}
 x_i\,\overline{x}_j\cdot \widetilde{s}~\text{fields}\;,
\end{equation}
where $x_i$ and $\overline{x}_j$ denote vector-like exotics. The terms
\eqref{eq:exotics} give rise to mass terms when the $\widetilde{s}$ acquire
VEVs, as discussed before in section~\ref{sec:CouplingsAndCo} and below
equation~\eqref{eq:exotics1}. It has been verified that the corresponding mass matrices have full
rank. Some of the relevant mass terms appear only at order 6 in the
$\widetilde{s}$ fields. As we have discussed earlier, we implicitly assume that
the $\widetilde{s}$ VEVs are not too large (in string units) so that we have
some sort of perturbative control. This raises the question whether  mass terms
arising at order 6 only are far below the GUT scale. This is not necessarily the
case: there are combinatorial factors in the superpotential couplings
\cite{Cvetic:1998gv} that partially undo the suppression coming from the high
powers in singlets. Further, the coefficients of the couplings have to be
calculated; recently general expressions for this purpose have been derived
\cite{Choi:2007nb}, but they have not yet been applied in concrete models.

In principle, one would expect that all vector-like states acquire large masses.
So one might lose the pair of Higgs doublets, which is also vector-like w.r.t.\
$G_\mathrm{SM}$. As we shall discuss now, this is not the case.
This issue is related to the GUT doublet-triplet splitting and the MSSM $\mu$
problems. In our construction the triplets sitting with the Higgs doublets
$h_{u,d}$ in the same multiplet get projected out by the orbifold action. Let us
now have a closer look at the effective $h_u\,h_d$ mass term. It turns out that
the pair $h_u\,h_d$ is neutral w.r.t.\ the selection rules, i.e.\ whenever a
coupling $h_u\,h_d\,\mathscr{M}$, with $\mathscr{M}$ denoting a monomial of
fields, $\mathscr{M}=\phi_1\dots\phi_n$, satisfies all selection rules, the
coupling $\mathscr{M}$ by itself represents an allowed coupling.  Further, it
was found \cite{Lebedev:2007hv} that (at least at order 6) the
global SUSY $F$-term equations are satisfied term by term,
\begin{equation}
 \frac{\partial\mathscr{M}}{\partial \widetilde{s}_i}~=~0\;,
\end{equation}
where $\mathscr{M}$ denotes a monomial of standard model singlets
$\widetilde{s}_i$ representing a superpotential term consistent with all
selection rules, this implies that also $\mathscr{M}$ vanishes. That is, in
supersymmetric vacua where the $F$-terms vanish term by term, all potential
superpotential terms are zero as well. Hence, at the perturbative level the
$h_u\,h_d$ mass term is zero. 

How can one obtain a $\mu$ term of the right size?
In the context of orbifold compactifications of the heterotic string, two
solutions to the $\mu$ problem have been proposed,
\begin{dingautolist}{'300}
 \item the proposal by Casas and Mu\~noz~\cite{Casas:1992mk} where $\mu$ originates
 from superpotential couplings to the hidden sector, and\label{CM}
 \item the proposal by Antoniadis et al.~\cite{Antoniadis:1994hg} where an
 effective $\mu$ term gets induced from the K\"ahler potential (for a recent,
 field-theoretic discussion along these lines see \cite{Hebecker:2008rk}).\label{A}
\end{dingautolist}
It turns out that both solutions can be employed in the model under
consideration (simultaneously). For the first solution \ref{CM}
\cite{Casas:1992mk}, one needs that the $\mu$ term be absent in supersymmetric
vacua and there has to be a relation between the $\mu$ term and the expectation value of
$\mathscr{W}$,
\begin{equation}
 \mu~\sim~\langle\mathscr{W}\rangle.
\end{equation}
The effective $\mu$ term is then generated by hidden sector dynamics, as in the
field-theoretic models by Kim et al.~\cite{Kim:1983dt,Chun:1991xm}.
Both criteria are met in the model under consideration. For solution \ref{A} to
work, the Higgs fields $h_{u,d}$ have to come from the untwisted sector
associated to a \Z2 twisted plane, as it happens to be the case in the model.
Then the K\"ahler potential has a favorable form,
similar to the one discussed in the Giudice-Masiero mechanism
\cite{Giudice:1988yz}. Altogether we see that all prerequisites for a successful
solution to the $\mu$ problem that have been discussed in the literature
\cite{Casas:1992mk,Antoniadis:1994hg} are present in the model. A more detailed
analysis of these issues is under way \cite{Brummer:2008pr}.

Let us now turn to the flavor structure of the model.
We are interested in the Yukawa couplings
\begin{equation}
 \mathscr{W}_\mathrm{Yukawa}
 ~=~
 Y_u^{fg}\,q_f\,\overline{u}_g\,h_u
 +
 Y_d^{fg}\,q_f\,\overline{d}_g\,h_d
 +
 Y_e^{fg}\,\ell_f\,\overline{e}_g\,h_d\;.
\end{equation}
At tree-level all Yukawa couplings vanish with the exception of the top
coupling, which is predicted to (roughly) coincide with the gauge coupling at
the high scale. All other Yukawas appear at higher order, i.e.\ are proportional
to different powers of $\widetilde{s}$ fields. Truncating at six singlets, one
obtains  
\begin{eqnarray}
Y_u&=& \left(
\begin{array}{ccc}
\widetilde{s}^5 & \widetilde{s}^5 &  \widetilde{s}^5 \\
\widetilde{s}^5 & \widetilde{s}^5 &  \widetilde{s}^6 \\
\widetilde{s}^6 & \widetilde{s}^6 &  1
\end{array}
\right)\;,\quad Y_d~=~ \left(
\begin{array}{ccc}
 0 & \widetilde{s}^5 & 0 \\
 \widetilde{s}^5 & 0 & 0 \\
 0 & \widetilde{s}^6 & 0
\end{array}
\right)\;,\quad Y_e~=~ \left(
\begin{array}{ccc}
 0 & \widetilde{s}^5 & \widetilde{s}^6 \\
 \widetilde{s}^5  & 0 & 0 \\
 \widetilde{s}^6 & \widetilde{s}^6 & 0
\end{array}
\right)\;.
\end{eqnarray}
As discussed, one should not take these textures literally. The $\widetilde{s}$
entries represent polynomials of a total of 46 different fields, which will have
hierarchies between their VEVs. Nevertheless it is remarkable that the Yukawas
exhibit the qualitatively right features:
\begin{dinglist}{"2B}
 \item $y_t \gg~$all other Yukawa couplings;
 \item hierarchies between the eigenvalues;
 \item non-trivial mixings. 
\end{dinglist}

 The models of the Mini-Landscape exhibit a non-Abelian discrete flavor symmetry,
$D_4$, for the two light generations \cite{Kobayashi:2004ya,Kobayashi:2006wq}.
This symmetry is only exact at the orbifold point, and broken in realistic
vacua. Nevertheless, using the $D_4$ symmetric situation as a starting point and
to then add corrections might have certain advantages when discussing the
(supersymmetric) flavor structure (cf.~\cite{Ko:2007dz}). The structure of the
soft masses is
\begin{equation}
 \widetilde{m}^2~=~\left(\begin{array}{ccc}
  a &  0 & 0\\
  0 &  a & 0\\ 
  0 &  0 & b
 \end{array}\right)+\text{terms proportional to $D_4$ breaking VEVs}\;.
\end{equation}
It is known that such an approximate form of the soft masses makes it possible
to avoid the supersymmetric flavor problems.

Let us now have a look at neutrino masses in this model.
In order to discuss neutrino masses in string-derived models one has first to
clarify what a (right-handed) neutrino is. In supersymmetric MSSM vacua
with $R$-parity this question is answered quite easily: a neutrino is an $R$-parity
odd $G_\mathrm{SM}$ singlet. In the model discussed so far, there are 49
neutrinos. Further, as discussed more generally in \cite{Buchmuller:2007zd}, all
ingredients of the see-saw \cite{Minkowski:1977sc} are present in this model
(already at order 6 in the standard model singlets):
\begin{itemize} 
 \item the right-handed neutrino mass matrix $M_\nu$ has full rank and
 \item neutrino Yukawa couplings $Y_\nu$ exist 
\end{itemize}
such that the effective neutrino mass matrix for the light neutrinos,
\begin{equation}
 m_\nu~=~v_u^2\cdot Y_\nu^T\,M^{-1}\,Y_\nu\;,
\end{equation}
has full rank.
This feature is again not specific to the model under discussion, we find more
generally that MSSM vacua of Mini-Landscape models have (typically, i.e.\ in all
models of the Mini-Landscape that we have analyzed so far) see-saw suppressed
neutrino masses. We also remark that, due to the large number of neutrinos, the
effective neutrino mass operator gets many contributions such that neutrino
masses are slightly enhanced against the naive estimate
$m_\nu^\mathrm{naive}\sim v_u^2/M_\mathrm{GUT}$. Let us also mention that
meanwhile the many neutrino scenario has been analyzed in some detail. It has
been found that many neutrinos somewhat relax the tension between leptogenesis
and supersymmetry \cite{Eisele:2007ws,Ellis:2007wz}. Furthermore, the presence
of many neutrinos and their Yukawa couplings have important consequences for
supersymmetric lepton flavor violation \cite{Ellis:2007wz}.

Because of the exact $R$-parity, dimension four proton decay operators are
absent in this model. However, it was found that both $q\,q\,q\,\ell$ and $\bar
u\,\bar u\,\bar d\,\bar e$ appear at order 6 in the SM singlets, and that they
are also generated by integrating out the heavy exotics. This leads to dimension
five proton decay, which is also a problem in 4D GUTs \cite{Dermisek:2000hr},
and might require further (perhaps discrete) symmetries \cite{Mohapatra:2007vd}.
There exists a large amount of (discrete) symmetries at the orbifold point.
The question whether letting some of them survive to low energies could suppress
proton decay is currently under investigation.
At this point, we are in the position to present a (probably biased) comparison
between the usual 4D GUTs and the string-derived higher-dimensional GUTs
(see table~\ref{tab:comparison}).
\begin{table}[h]
\begin{tabular}{l|p{5cm}|p{5cm}}
framework & 4D SO(10) GUTs & `local GUTs'\\
\hline
spectrum & $3\times\boldsymbol{16}+\boldsymbol{10}$ & 
$3\times\text{generation}+\text{Higgses}$\\
$\phantom{\text{masses of extra particles}}$ & $+\boldsymbol{210}+\overline{\boldsymbol{126}}
+\dots$
& $+\text{vector-like exotics}$
\\
& 
\multicolumn{2}{c}{$\underbrace{\hspace*{10cm}}_{\displaystyle\text{few}\times100\text{ degrees of
freedom}}$}

\\
\hline
masses of extra particles & $\sim M_\mathrm{GUT}$ & $\sim M_\mathrm{string}\times
\left(\frac{\text{singlet VEV}}{M_\mathrm{string}}\right)^n$ 
\\
\hline
doublet triplet splitting & 
cancellation %\\
between different %\\
vacuum expectation %\\
values necessary
&
automatic %\\
connection %\\
$\mu\leftrightarrow\langle\mathscr{W}\rangle=m_{3/2}$
\\
\hline
GUT symmetry breaking & appropriate Higgs fields need to be introduced 
& encoded in geometry \& gauge embedding (choice `by hand')
\\
\hline
gauge coupling unification & \checkmark &  \checkmark
\\
\hline
see-saw & \checkmark & \checkmark
\\
\hline
GUT relations for masses & \checkmark & $\boldsymbol{\times}$
\\
\hline
$R$- or matter parity & \checkmark & \checkmark
\\
\hline
proton stability & requires additional symmetries & 
problem of dimension five operators not yet solved
\end{tabular}
\caption{Comparison between 4D and string-derived higher-dimensional GUTs.}
\label{tab:comparison}
\end{table}

Let us now answer the questions raised at the beginning of the section. First of
all, we can indeed qualitatively reproduce the MSSM with all necessary features
(matter content, gauge interactions with the correct coupling strengths,
hierarchical Yukawa couplings with non-trivial mixing, see-saw suppressed
neutrino masses, a hidden sector dynamically explaining soft masses in the right
range, etc.). However, we cannot claim to be able to compute the precise
numerical values of standard model parameters (such as the electron
mass).\footnote{The
obstacles to such a computation are well known: one would have to come up with a
theory of moduli stabilization that allows to precisely predict the values of
the moduli. With these one could compute the superpotential couplings. Then one
might analyze the (local) minima of the scalar potential for the $\widetilde{s}$
fields, thus determining their VEVs. (It is quite possible that there is an
interplay between fixing the geometric moduli and minimizing the scalar
potential of the $\widetilde{s}$ fields, which further complicates the issue.)
These $\widetilde{s}$ VEVs need to be inserted in order to compute the Yukawa
couplings. Taking (quantum) corrections to the K\"ahler potential into account,
one could then obtain an estimate on the physical couplings. As discussed
before, this would have to be done for a large number of vacua, which we cannot
even count at the moment. That is, there is still a very long way to go, and
deriving models with exact MSSM spectrum and qualitatively right features
represents only the first steps of this Marathon program.}
On the other hand, we have seen that string theory gives us more than just
reproducing (known) features. For instance, the fact that the $\mu$ problem is
solved automatically is very encouraging, and might not just be an accident. 
We will have yet to see whether similar statements can be made with regards to
other puzzles such as the strong CP problem.

\subsection{MSSM-like models from the $\boldsymbol{\Z{12}}$-I orbifolds}

So far we have focused on \Z6-II orbifolds. It turns out that in the \Z{12}-I
orbifold geometry promising models can be found as well. The exploration of this
geometry started with \cite{Kim:2006hv,Kim:2006hw}, where models with the
flipped \SU5 gauge group, $G_\mathrm{fl}=\SU5'\times\U1_X$, were constructed.
The advantage of this gauge symmetry is that one does not need large
representations in order to break $G_\mathrm{fl}$ to $G_\mathrm{SM}$; it is
sufficient to let an \SU5 $\boldsymbol{10}$ acquire a VEV. 
The question of $R$-parity has been analyzed in this class of models as well 
\cite{Kim:2006zw}. It was found that one can either have an exact $R$-parity and
states beyond the MSSM matter content or remove the exotics and have only an
approximate $R$-parity, which might not necessarily be inconsistent with
observation. The question of getting a phenomenologically viable axion from
these constructions has also been addressed~\cite{Choi:2006qj}.
Another model with an effective $R$-parity, but not based on flipped \SU5, has
been presented in \cite{Kim:2007mt}.

\subsection{Comments on the SO(32) heterotic string}

Early works on orbifold compactifications mainly focused on the $\E8\times\E8$
heterotic string, leaving the \SO{32} version, to a large extent, unexplored.
One of the reasons for this is that the coset $\SO{32}/\SO{10}$ does not contain
$\boldsymbol{16}$-plets, implying that it is impossible to obtain standard model
families from the untwisted sector. Further, the simplest, i.e.\ $\Z3$, orbifold
does not have any \SO{10} spinors in its massless spectrum at all. This lead to
the expectation that it is difficult to get phenomenologically viable models
from the \SO{32} theory.

Only recently, the interest in four-dimensional heterotic \SO{32} orbifold
constructions has been revived~\cite{Giedt:2003an,Nibbelink:2007rd}. In the case
of \Z{N} orbifolds, a complete classification of gauge embeddings in absence of
background fields has been obtained~\cite{Choi:2004wn,Nilles:2006np}.
Interestingly, that classification has shown that spinors of \SO{2n} gauge
groups appear rather frequently in the twisted sectors of \SO{32} orbifolds. 
Spinors of \SO{10}, in particular, are found locally at fixed points of the
first twisted sector of many orbifold models.  Moreover, that classification
reveals that the amount of available \SO{32} orbifold models is comparable, at
the same level, to that of its more famous brother, the $\E8\times\E8$ string.
From this we conclude that model building based on the
heterotic \SO{32} string theory might be as interesting as that based on the
$\E8\times\E8$ theory.

Let us also remark that the appearance of spinors in models derived from the \SO{32}
heterotic string will also be important for completing the understanding of the
\SO{32} heterotic type I duality in four space-time dimensions. We know that
spinors do not appear in the perturbative type I theory.

\section{Orbifold GUT limits}

Much of the success of the orbifold GUTs is due to the fact that they provide a
very simple understanding of the power of gauge symmetry breaking in extra
dimensions. Here we discuss a top-down motivation for the orbifold GUT scheme,
following \cite{Hebecker:2004ce}. It is well known that there is a (rather
small) discrepancy between the scale where the gauge couplings meet,
$M_\mathrm{GUT}$, and the string scale $M_\mathrm{string}~=~(\alpha')^{-1/2}$. 
The heterotic string scale can be defined as the ratio of 10D gauge and
gravitational coupling strength, $m_\mathrm{het}=\frac{g_{10}}{\kappa_{10}}$ in
standard notation (cf.\ e.g.\ \cite{Hebecker:2004ce}). It can be interpreted as
the mass of the lowest-lying massive string state. The 10D and 4D gauge
couplings are related by
\begin{equation}
 g_4^2~=~\frac{g_{10}^2}{\mathcal{V}}\;,
\end{equation}
where $\mathcal{V}$ denotes the volume of the 6D compact space. On the other
hand, the string scale in the heterotic case is fixed, 
using $\alpha= g_4^2/4\pi\simeq1/25$ one has
\begin{equation}
 M_\mathrm{string}~=~(\alpha')^{-1/2}~\simeq~8.6\cdot10^{17}\,\mathrm{GeV}
 \quad\text{while}\quad
 M_\mathrm{GUT}~\simeq~(2-3)\cdot10^{16}\,\mathrm{GeV}\;,
\end{equation} 
i.e.\ there is a discrepancy
\begin{equation}\label{eq:discrepancy}
 \frac{M_\mathrm{string}}{M_\mathrm{GUT}}~\sim~30\;.
\end{equation}
We regard it as an encouraging fact that both scales are rather close. This
gives further credit to the `grand desert' picture of MSSM gauge coupling
unification; in other schemes one typically has to work much harder to obtain
similar agreement. Nevertheless, the  discrepancy \eqref{eq:discrepancy} asks
for an explanation. The probably most far-reaching proposal is that of
Ho{\v{r}}ava and Witten~\cite{Horava:1995qa,Horava:1996ma}, which we, however,
do not discuss in detail here. In a footnote, Witten has pointed out a very simple
solution: the GUT scale \emph{can} be related to a compactification scale if
internal space is anisotropic~\cite[footnote~3]{Witten:1996mz}. 
For an isotropic toroidal compactification one has
\begin{equation}
 \frac{\alpha_\mathrm{GUT}}{2}~=~\frac{g_\mathrm{het}^2}{(R\,m_\mathrm{het})^6}
 \;.
\end{equation}
Anisotropic compactification may mitigate the discrepancy between $M_\mathrm{GUT}$ and $M_\mathrm{string}$
\cite{Witten:1996mz,Hebecker:2004ce}. 
For example, if one assumes a toroidal compactification and chooses one radius
much larger than the other 5, one has
\begin{equation}
 \frac{\alpha_\mathrm{GUT}}{2}
 ~=~
 \frac{g_\mathrm{het}^2}{(R_\mathrm{large}\,m_\mathrm{het})\,(R_\mathrm{small}\,m_\mathrm{het})^5}
 \;.
\end{equation}
This means that for $g_\mathrm{het}\simeq1$ and $R_\mathrm{small}\simeq
m_\mathrm{het}^{-1}$ one can choose $R_\mathrm{large}\,m_\mathrm{het}\sim50$ or
$R_\mathrm{large}^{-1}\sim 3\cdot10^{16}\,\mathrm{GeV}$, i.e.\ the large radius
can be of the order of the GUT scale. Although there might be some obstructions
for taking just one radius much larger than the other
\cite[footnote~7]{Hebecker:2004ce}, this result indicates that taking one or two
radii much larger than the others can mitigate the discrepancy between
$M_\mathrm{GUT}$ and $M_\mathrm{string}$.
As discussed in \cite{Witten:1996mz}, the amelioration resulting from the
anisotropy might have to be combined with the strong coupling effect
described in~\cite{Horava:1995qa,Witten:1996mz,Horava:1996ma}.

Let us remark that recently it has been shown that, using  Casimir
stabilization, a radius $R\sim M_\mathrm{GUT}^{-1}$ appears possible
\cite{Buchmuller:2008cf}. Of course, the question why only one radius gets
stabilized by this mechanism while the others are smaller by a factor $10$ or
more needs still to be answered.

\begin{figure}[!h!]
\begin{center}
\CenterEps{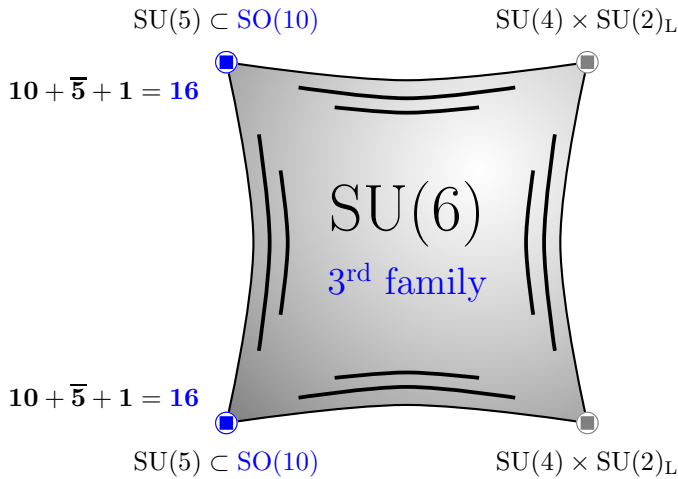}
\end{center}
\caption{Orbifold GUT limit of model 1A of \cite{Lebedev:2007hv}.}	
\label{fig:OGL}	
\end{figure}

This leads then to the following picture: the world is four-dimensional up to
the scale, where one or two extra dimensions open up and gauge couplings meet.
To be specific, let us discuss one particular orbifold GUT limit of the model
discussed in \ref{sec:BenchmarkModel}. Assuming that the radii of the \SO4 torus
be large, one arrives at an orbifold GUT where two families are localized at
\SU{5} fixed points while the third family comes from the bulk
(figure~\ref{fig:OGL}).  In the limit in which the vertical direction gets
small, one arrives at a setting that shares many features with the 5D orbifold
GUT presented in \cite{Burdman:2002se}. An important feature of this setting is
that the bulk group $\SU6$ contains the standard model as a subgroup,
$G_\mathrm{bulk}\supset G_\mathrm{SM}$. Since the running above the
compactification scale is dominated by the bulk, radiative corrections will be
universal, i.e.\ not discriminate between the standard model gauge factors.

\begin{figure}[h]
\centerline{\includegraphics{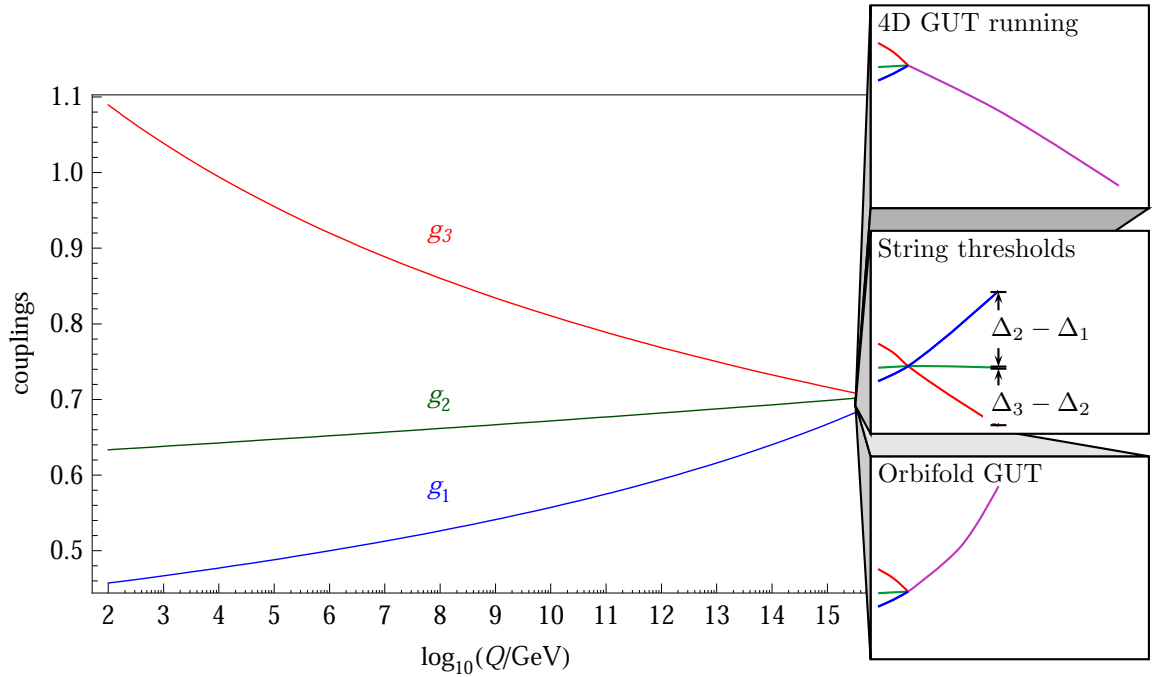}}
\caption{Different possibilities for gauge coupling unification.}
\label{fig:GCV}
\end{figure}

Let us now compare different possibilities for gauge coupling unification
(figure~\ref{fig:GCV}). The simplest possibility is that of a 4D GUT, where
a unified running occurs above $M_\mathrm{GUT}$ (upper inlay in
figure~\ref{fig:GCV}). However, this picture is challenged by the fact that the
Higgs fields in large representations, which are needed to break the GUT group,
are hard to obtain. Another possibility, that resolves the discrepancy between
string and GUT scales, has been discussed in the 90ies
\cite{Ibanez:1991zv,Nilles:1995kb,Nilles:1997vk}. 
There is a sort of mirage unification, i.e.\ there is no physical significance
to $M_\mathrm{GUT}$. The theory has still the gauge symmetry of the MSSM almost
up to the string scale, where the six extra dimensions (more or less)
simultaneously open up. Gauge couplings of the different gauge factors keep on
running until this scale. Above the compactification scale, running in extra
dimensions takes place, which can be described in terms of string threshold
corrections $\Delta_i$. These thresholds are different for different factors of
$G_\mathrm{SM}$, and can account for the mismatch between the couplings (middle
inlay in figure~\ref{fig:GCV}). However, in this picture the unification of the
three gauge couplings in a point appears somewhat miraculous. The third
possibility is the one of the orbifold GUT picture, where at $M_\mathrm{GUT}$
one or two extra dimensions open up. Above $M_\mathrm{GUT}$ the running is
dominated by the bulk group, which unifies the standard model gauge factors,
$G_\mathrm{bulk}\supset G_\mathrm{SM}$ (see also \cite{Dundee:2008ts}). The
running is hence universal (lower inlay in figure~\ref{fig:GCV}).\footnote{It
has been pointed out that there might be non-universal contributions from
localized fields~\cite{Hebecker:2004ce}. The resulting corrections are, however,
small; they are comparable to supersymmetric thresholds at the TeV scale which
might arise because squarks are heavier than sleptons and so on. Both
corrections have to be taken into account (as well as contributions from
vector-like exotics), but they do not necessarily destroy the beautiful picture
of MSSM gauge coupling unification. }  This possibility appears particularly
attractive and deserves to be studied in more detail.

\section{Orbifolds vs.\ Calabi-Yau compactifications}
\label{sec:BlowUp}

We have discussed how one can construct MSSM-like models by compactifying on an
orbifold and then selecting supersymmetric vacua in which certain (SM singlet)
fields attain vacuum expectation values. It is known that giving VEVs to fields 
localized at orbifold singularities corresponds to blowing up the orbifold,
i.e.\ resolving the orbifold singularities
\cite{Dixon:1986jc,Hamidi:1986vh,Dixon:1986qv,Walton:1988bu,Erler:1994zy,Aspinwall:1994ev}.
After the blow-up procedure one obtains a Calabi-Yau (CY) manifold  (or at least
something that looks in some regions like a CY). For instance, blowing up the
\Z3 orbifold is known to yield the standard
CY~\cite{Dixon:1986jc,Hamidi:1986vh,Dixon:1986qv}. In the process of blowing up,
\begin{itemize}
 \item the gauge symmetry gets reduced;
 \item pairs  of states which are vector-like w.r.t.\ to the remnant gauge
 symmetry may get massive.
\end{itemize}
In the CY picture, one does usually not discuss the masses of the states which
pair up away from the orbifold point. In a \emph{generic} point of the moduli
space these masses will be of the order of the heterotic string scale. On the
other hand, if the blow-up fields are moduli, as often happens to be the case,
then there is a smooth interpolation between the orbifold point and a generic
CY vacuum. The masses will then depend on `how generic' the point in moduli
space is. As long as the VEVs of the blow-up moduli are small, one retains
perturbative control over the setting, and can calculate the masses of the
fields as well as other features of the model. On the other hand, there are
features that do not depend at which point in moduli space one is sitting; the
chiral spectrum of the model being one example.

The blow-up procedure, i.e.\ the interpolation between the orbifold point and a
CY point has been revisited recently in a series of
papers~\cite{Honecker:2006qz,%
GrootNibbelink:2007ew,GrootNibbelink:2007pn,GrootNibbelink:2007rd,Nibbelink:2008tv}
(see \cite{Lust:2006zh} for related work). The focus of these works is the
resolution of local singularities $\mathbbm{C}^n/\mathbbm{Z}_n$, especially
$\mathbbm{C}^3/\mathbbm{Z}_3$. Very much the same as in the local GUT scheme,
one starts with the compact orbifold, say $\mathbbm{T}^6/\mathbbm{Z}_3$, and
zooms into a given fixed point such that the local model
$\mathbbm{C}^n/\mathbbm{Z}_n$ is described by the local shift $V_\text{local}$
only. This local shift determines the local gauge group and the local untwisted
and twisted matter spectrum. Then, the blow-up is induced by a VEV of a single
twisted field, the blow-up mode. In order to see the blow-up procedure in
detail, the corresponding smooth resolution space $\mathcal{M}^3$ of
$\mathbbm{C}^n/\mathbbm{Z}_n$ is constructed explicitly. $\mathcal{M}^3$ is
parametrized by a parameter $r$ defining the radius of the resolved singularity,
i.e.\ when $r\rightarrow 0$ the resolved space approaches the singular orbifold
$\mathbbm{C}^3/\mathbbm{Z}_3$. In order to compactify the heterotic string on
$\mathcal{M}^3$ a specific $\U{1}$ bundle corresponding to the local shift
$V_\text{local}$ is chosen. This choice completely defines the resolved model.
Note, however, that the blow-up of an orbifold is \emph{not unique}.
Rather, as already remarked in \cite{Dixon:1986jc}, the orbifold point might be
thought of as a junction in moduli space that can be blown up to different
smooth spaces.
However, the blown-up orbifold model and the resolved model seem to differ in
general: for example, there can be two anomalous $\U{1}$'s on the resolution in
contrast to a single anomalous $\U{1}$ on the orbifold. Nevertheless, these
discrepancies can be resolved when one carefully analyzes the role of the
blow--up mode: on the orbifold side, the standard Green-Schwarz mechanism,
involving one single universal axion, is combined with a Higgs mechanism giving
rise to the blow--up. On the resolution, this combination is mapped into a
Green--Schwarz mechanism involving two axions. These axions are mixtures of the
orbifold axion and of the blow--up mode. Therefore, one can achieve a complete
matching between the blown--up local orbifold models and the corresponding
smooth resolutions. Finally, in \cite{Nibbelink:2008tv} it is shown how the
local blow--up procedure can be applied to the compact case
$\mathbbm{T}^6/\mathbbm{Z}_3$, even in the presence of Wilson lines. As an
application, the blow--up of an early MSSM candidate constructed from
$\mathbbm{T}^6/\mathbbm{Z}_3$ with "3 generations +
vector-like"~\cite{Ibanez:1987sn} is analyzed. It turns out that a full blow-up
of all fixed points would destroy the nice properties of this model, e.g.\ it
would break the hypercharge near the string scale. On the other hand, a partial
blow-up of some fixed points can retain the MSSM properties. Additionally,
since many vector-like exotics have trilinear couplings to the blow-up modes,
they become heavy yielding a natural explanation for the decoupling of the
vector-like matter. The main lessons for phenomenology from this discussion
might be that reasonable particle physics models can be regarded as/obtained
from partial blow-ups of orbifolds.

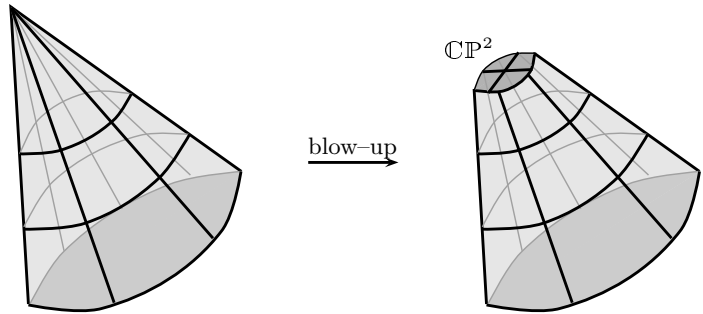
\begin{figure}[h]
\begin{center}
\psset{unit=0.6cm}
\begin{pspicture}(3cm,3cm)(13cm,8cm)
%%
%% Depth: 2147483647
%%
\newgray{mycolor0}{0.63}\definecolor{mycolor0}{gray}{0.63}
%%
%% Depth: 50
%%
\psset{linestyle=solid,linewidth=0.0635,linecolor=black,fillstyle=solid,fillcolor=gray90}
\psline(5.80,6.33)(5.41,12.92)(10.52,9.27)
\psline(15.95,6.33)(15.57,12.92)(20.67,9.27)
\psset{linecolor=black,fillstyle=none}
\psline{->}(12.00,9.47)(14,9.47)
\rput[l](12.00,9.78){blow--up}
%%
%% Depth: 40
%%
\psset{linewidth=0.03175,linecolor=mycolor0,fillstyle=solid,fillcolor=gray80}
\pscurve(20.68,9.27)(19.09,8.99)(16.71,7.47)(15.97,6.30)
\psset{linewidth=0.0635,linecolor=black}
\pscurve(15.95,6.31)(17.57,6.23)(20.24,7.95)(20.67,9.27)
\psset{linewidth=0.03175,linecolor=mycolor0}
\pscurve(10.53,9.27)(8.94,8.99)(6.56,7.47)(5.82,6.30)
\psset{linewidth=0.0635,linecolor=black}
\pscurve(5.80,6.31)(7.42,6.23)(10.08,7.95)(10.51,9.27)
%%
%% Depth: 10
%%
\psset{linewidth=0.03175,linecolor=mycolor0,fillstyle=none}
\psline(6.59,7.51)(5.42,12.90)
\psline(9.39,9.19)(5.42,12.90)
\psline(7.88,8.44)(5.42,12.90)
\psline(19.54,9.19)(15.57,12.90)
\psline(16.74,7.51)(15.57,12.90)
\psline(18.04,8.44)(15.57,12.90)
\pscurve(15.86,8.04)(16.51,9.06)(18.42,10.07)(19.51,10.09)
\pscurve(15.76,9.66)(16.19,10.44)(17.32,11.03)(18.24,10.99)
\pscurve(5.70,8.04)(6.36,9.06)(8.27,10.07)(9.36,10.09)
\pscurve(5.61,9.66)(6.04,10.44)(7.16,11.03)(8.08,10.99)
%%
%% Depth: 5
%%
\psset{linewidth=0.0635,linecolor=black}
\psline(5.42,12.90)(7.70,6.36)
\psline(5.42,12.90)(9.89,7.79)
\psline(15.57,12.90)(17.85,6.36)
\psset{fillstyle=solid}
\psline(15.57,12.90)(20.05,7.79)
\psset{fillstyle=none}
\pscurve(5.61,9.66)(6.51,9.74)(7.70,10.42)(8.08,10.99)
\pscurve(5.71,7.99)(7.05,8.12)(8.70,9.13)(9.37,10.09)
%%
%% Depth: 2
%%
\psset{linewidth=0.03175,linecolor=white,fillstyle=solid,fillcolor=gray70}
\pspolygon*(15.50,11.08)(16.01,11.14)(16.95,11.81)(17.20,12.04)(15.75,13.11)(15.34,12.92)(15.50,11.08)
\psset{linewidth=0.0635,linecolor=black,fillcolor=gray70}
\pscurve(15.67,11.07)(16.34,11.06)(16.91,11.47)(17.02,11.91)
\psset{linewidth=0.03175,linecolor=black}
\pscurve(15.68,11.07)(15.88,11.51)(16.47,11.86)(17.03,11.89)
%%
%% Depth: 1
%%
\psset{linecolor=mycolor0,fillstyle=none}
\psline(16.65,11.89)(16.95,11.62)
\psline(16.22,11.73)(16.52,11.18)
\psline(15.88,11.50)(15.98,11.04)
\psset{linewidth=0.0635,linecolor=black}
\psline(15.99,11.01)(16.66,11.91)
\psline(16.97,11.53)(15.86,11.48)
\psset{linecolor=black}
\pscurve(15.76,9.66)(16.66,9.74)(17.86,10.42)(18.24,10.99)
\pscurve(15.86,7.99)(17.20,8.12)(18.85,9.13)(19.52,10.09)
\rput[l](15,12){$\mathbbm{CP}^2$}
\end{pspicture}\psset{unit=1cm}
%% End
\end{center}
\caption{Blowing up orbifold singularities.}
\label{fig:BlowUp}
\end{figure}

Now one might ask why not to compactify directly on a CY rather than starting
with the orbifold point and then moving away in moduli space. Indeed, CY
compactifications are known to lead to promising models
\cite{Pokorski:1998hr,Donagi:1999ez,Donagi:2004su,Donagi:2004ub,Braun:2005ux,Bouchard:2005ag,Braun:2006ae}
(GUT models were derived from \U{N} line bundles
\cite{Blumenhagen:2006ux,Blumenhagen:2006wj}).
It is, in particular, reassuring that models with the exact MSSM spectrum have
been found in this context as well.

The couplings in a potentially realistic CY compactification have been
calculated and discussed \cite{Bouchard:2006dn}. The qualitative picture is as
follows. The Yukawa coupling is either order one or zero, however, it appears
difficult to obtain hierarchical couplings at a \emph{generic} point in moduli
space. This does not mean that the CY is incapable of reproducing phenomenology,
but it might mean that one would have to move close to a special point in moduli
space. Whether this point can, at least to some extent, be identified with a
configuration close to an orbifold point (or a partial blow-up of an orbifold)
represents an interesting question. In fact, in the CY model
\cite{Bouchard:2005ag,Bouchard:2006dn} there is no light Higgs pair at a generic
point in moduli space: the $\mu$ term only  vanishes if the size of a certain
cycle shrinks to zero. A $\mu$ term of order TeV forces us to move to a  rather
special point in moduli space.

Altogether we see that neither the orbifold point nor the generic CY
configuration in moduli space appear to describe observation. We have argued
that vacua not too far from the orbifold point exhibit various
phenomenologically desirable properties (but, of course, there might be
different possibilities). Currently different strategies to explore the
phenomenologically interesting region in moduli space are pursued: one can
either start from the orbifold point or directly compactify on CY spaces. It
might well be that only a combination of both strategies will be ultimately
successful.

\section{Summary}

We have described how gauge symmetry breaking in extra dimensions allows to
resolve some of the most vexing problems in 4D grand unification. 
We discussed how orbifold GUTs can be derived from heterotic string theory.
The resulting constructions have a very simple geometric interpretation. 
The key aspects of gauge group topography and local grand unification
have been described in some detail. With these guidelines it is possible to 
construct string models with the exact MSSM matter content and $R$-parity. Because 
string-derived, these models have certain additional (and surprising) relations 
which are, as opposed to  field-theoretic constructions, not `put in by hand'. The 
most striking features (or `stringy surprises') are:
\begin{dinglist}{"50}
 \item a statistical preference for low-energy supersymmetry;
 \item a relation between the $\mu$ term and the gravitino mass;
 \item see-saw for free;
 \item $y_t\gg~\text{other Yukawa couplings}$.
\end{dinglist}
It is very tempting to believe that these features are not just accidents, but
somehow tell us that we are not on a completely wrong track. 

We have derived models from string theory which are not immediately ruled out by
observation (e.g.\ due to the prediction of unobserved states). On the other
hand, we cannot claim that we have found `the' MSSM (nor do we know that all our
constructions are incapable to reproduce `the' MSSM). To obtain quantitative
predictions from our constructions (such as on the electron mass) one would have
to explore the vacua of the models in much greater detail. As of now, not all
necessary tools are available, so it will require major effort to accomplish
this task. But given the progress over the past few years we feel that this
effort will be justified.

\subsection*{Acknowledgments}

We would like to thank Stephan Stieberger for discussions. This research was
supported by the DFG cluster of excellence Origin and Structure of the Universe,
the European Union 6th framework program MRTN-CT-2004-503069 "Quest for
unification", MRTN-CT-2004-005104 "ForcesUniverse", MRTN-CT-2006-035863
"UniverseNet" and SFB-Transregios 27 "Neutrinos and Beyond" and 33 "The Dark
Universe" by Deutsche Forschungsgemeinschaft (DFG).

\bibliography{Orbifold}
\bibliographystyle{epj}

\end{document}